\documentclass[ejs, submission]{imsart}
\RequirePackage[numbers]{natbib}
\RequirePackage[OT1]{fontenc}
\usepackage[utf8]{inputenc}
\setcitestyle{aysep={}}
\RequirePackage{hyperref}
\usepackage{amsmath}
\usepackage{amssymb}
\usepackage{graphicx}
\usepackage{multicol}
\usepackage{xcolor}
\usepackage{comment}
\usepackage{amsfonts}
\usepackage{float}
\usepackage{outlines}
\usepackage[ruled, vlined]{algorithm2e}
\usepackage{dsfont}
\usepackage[thinlines]{easytable}
\usepackage{subcaption}
\usepackage{amsthm}
\newtheorem{theorem}{Theorem}[section]
\newtheorem{lemma}{Lemma}[section]

\newtheorem{definition}{Definition}[section]
\usepackage{enumitem}

\DeclareMathOperator*{\argmin}{arg\,min}
\DeclareMathOperator*{\minimize}{minimize}

\newcommand{\Ell}{\mathcal{L}}
\newcommand{\hide}[1]{}

\DeclareMathOperator{\bbeta}{\boldsymbol \beta}
\def\X{\mathbf X}
\def\Y{\mathbf Y}
\def\Z{\mathbf Z}
\def\y{\mathbf y}
\def\bxi{\mathbf x_i}
\def\Ell{\mathcal L}
\def\Q{\mathbf Q}
\numberwithin{equation}{section}
\theoremstyle{plain}
\newtheorem{corollary}{Corollary}

\begin{document}

\begin{frontmatter}
\title{Sparse Regression for Extreme Values}
\runtitle{Sparse Regression for Extreme Values}

\begin{aug}
\author{\fnms{Andersen} \snm{Chang}\ead[label=e1]{atc7@rice.edu}}
\and
\author{\fnms{Minjie} \snm{Wang}\ead[label=e2]{mw50@rice.edu}}

\address{Department of Statistics, Rice University\\
\printead{e1,e2}}

\author{\fnms{Genevera I.} \snm{Allen}
\ead[label=e3]{gallen@rice.edu}}

\address{Department of Electrical and Computer Engineering, Rice University,\\
Department of Computer Science, Rice University,\\
Department of Statistics, Rice University,\\
Department of Pediatrics-Neurology, Baylor College of Medicine,\\
Jan and Dan Duncan Neurological Research Institute, Texas Children’s Hospital \\
\printead{e3}\\}

\runauthor{A. Chang, M. Wang, and G. I. Allen}

\end{aug}
  
\begin{abstract}
    We study the problem of selecting features associated with extreme values in high dimensional linear regression. Normally, in linear modeling problems, the presence of abnormal extreme values or outliers is considered an anomaly which should either be removed from the data or remedied using robust regression methods. In many situations, however, the extreme values in regression modeling are not outliers but rather the signals of interest; consider traces from spiking neurons, volatility in finance, or extreme events in climate science, for example. In this paper, we propose a new method for sparse high-dimensional linear regression for extreme values which is motivated by the Subbotin, or generalized normal distribution, which we call the extreme value linear regression model. For our method, we utilize an $\ell_p$ norm loss where $p$ is an even integer greater than two; we demonstrate that this loss increases the weight on extreme values. We prove consistency and variable selection consistency for the extreme value linear regression with a Lasso penalty, which we term the Extreme Lasso, and we also analyze the theoretical impact of extreme value observations on the model parameter estimates using the concept of influence functions. Through simulation studies and a real-world data example, we show that the Extreme Lasso outperforms other methods currently used in the literature for selecting features of interest associated with extreme values in high-dimensional regression. 
    
\end{abstract}

\begin{keyword}[class=MSC]
\kwd[Primary ]{62J05}
\kwd{62J07}
\kwd[; secondary ]{62P10}
\kwd{62P05}
\end{keyword}

\begin{keyword}
\kwd{linear regression}
\kwd{sparse modeling}
\kwd{extreme values}
\kwd{Subbotin distribution}
\kwd{generalized normal distribution}
\end{keyword}
\tableofcontents
\end{frontmatter}

\newpage

\section{Introduction}

When applying linear regression models, one often encountered issue is the presence of rare extreme values, defined here as abnormally large magnitude observations. This can occur in the form of outliers in the response variable as well as in the form of highly influential points in the predictor variables. Historically, statisticians have tried to develop methods to ignore or dampen the effects of outliers in data sets when doing a linear regression analysis. Metrics such as residual analysis, Cook's distance, and DFFIT can be used to identify and possibly remove outliers from the data set \citep{rob}. New regression methods have also been developed to handle outliers in response variables as well. For example, robust regression \citep{rob2} has been used in many different applications, and much work has been to done to show theoretical asymptotic performance in the presence of outliers \citep{rob4, rob5}. More recently, several have studied robust regression procedures for high-dimensional data \citep{loh2017statistical, Eunhoyang}.

However, in certain contexts, the important information in the response variable that we want to model or predict is in the rare, abnormally large magnitude observations. For these types of applications, rather than wanting to remove outliers or use robust regression methods, we instead want to focus on these extreme values when fitting models to the data. For example, in neuroscience, calcium imaging data collected contains measurements of fluorescence traces of neurons in the imaged brain \citep{cal}; the signal that is important in this situation is the occurrences of neuron firing, indicated by large positive spikes in the fluorescence trace. Extreme value regression models are often used as well in climatology to measure the rate and strength of extreme climate or weather events \citep{ev2}, or in finance to predict periods of high volatility of asset prices \citep{ev1}. Their potential usage has also been studied in spectroscopy analysis and signal processing\citep{ev9}.

Several different possible approaches to the problem of high-dimensional regression for extreme values have been used in various fields. Sparse regression methods based on classical extreme value theory utilize a generalized linear model framework. The extreme values above a predetermined threshold in a response variable are specified to follow a distribution, such as the Gumbel, whose parameters are a linear function of the predictor variables and which determine the frequency and magnitude of the extreme values \citep{evr,evr2}. Another regression model commonly applied to model extreme values in the high-dimensional setting is sparse quantile regression, specifically applied to a very high or very low quantile \citep{koe}. These types of models use a weighted absolute deviation loss function in order to find the expected value of a response variable at a particular quantile. Extensions to high-dimensional sparse $\ell_1$ quantile regression have also been studied extensively \citep{belloni, li}. These types of regression methods have shown to be effective for finding features which are correlated to larger magnitude values of a response variable when there is ample data to create a reliable model. In the types of applications we are considering, though, the extreme values tend to be very rare for a typical set of observations. Because of this, it is unclear how the desired quantile should be chosen based on the number and magnitude of the extreme events. The rarity of the extreme values can also cause the results from the regression model to be numerical unstable due to the lack of adequate data to get accurate estimates and to sensitivity to choice of quantile at the extremes. Additionally, quantile regression will not be as useful in the situation when the response variable of interest has both positive and negative extreme values, as the model by construction will upweight the impact of one side of the extreme values while heavily downweighting the other. Thus, quantile regression potentially restricts us to focusing only on some of the extreme values while essentially ignoring others.

One other widely-used approach for modeling extreme values involves pre-processing the data via some type of thresholding algorithm, keeping only the observed values of each variable which are above either a static or dynamic threshold and zeroing out the others. Examples of this in different fields include spike calling or deconvolution in neuroscience \citep{theis} or Otsu's method in image processing \citep{otsu}. After these algorithms have been applied to the data, typical high-dimensional regression methods are then applied to the data. In general, thresholding data can help in regression analysis for extreme values by removing any influence from non-extreme values. However, this type of filtering is not necessarily desirable in all situations. Thresholding approaches by their nature binarize the observations of a variable in to extreme and non-extreme categories, whereas in some cases it may make more sense to smooth the transition from extreme to non-extreme values if it is not clear where the boundary between the two should lie. Also, the addition of an extra data pre-processing step can potentially lead to less precise estimates from the following regression analysis, since any errors made in the former will propagate to the latter regression step. 

In this paper, we explore a different potential approach to tackle the problem of modeling and predicting extreme values. Our approach to this problem is to increase the relative weight of larger magnitude losses compared to regular ordinary least squares. Conceptually, this problem is analogous to increasing the power of the Gaussian kernel function, which leads to the generalized normal distribution \citep{sub}. Thus, we base our method on $\ell_p$ norm regression, which uses a general $p$ norm for regression rather than the ordinary $\ell_2$ norm.  This is a method which has been well-studied as a whole in the past in the statistics literature \citep{lp1, lp2}. However, much of the effort in previous research has been focused on showing that $\ell_p$-norm regression can be more robust to outliers \citep{lp4,lp5} by using a norm between 0 and 1. On the other hand, we are interested in using this type of regression model to create a method which is more sensitive to extreme values in the response by using norms larger than the squared error loss, i.e. when $p > 2$. By doing this, we skew the regression results toward finding the relationships with extreme values in a response variable without disregarding potentially useful observations that could otherwise be ignored by thresholding or substantially downweighted by quantile regression. We also analyze the theoretical influence of extreme value observations on our proposed regression model as well as the finite sample performance guarantees of the estimation procedure. While general theoretical properties of $\ell_p$ norm regression have been examined in previous literature \citep{lp3}, the performance with respect to regression for extreme values when $p > 2$ for $\ell_p$ norm regression has not been well-studied; this particular situation presents its own unique theoretical and practical challenges, which we will investigate in this paper.

The rest of the paper is organized as follows. Section 2 introduces and characterizes the extreme value linear regression method and presents the algorithm used for parameter estimation. We then prove consistency and sparsistency results in section 3. Lastly, in section 4, we investigate the performance of the extreme value linear regression through simulation and real data studies.

\section{Regression for Extreme Values}

Let $\mathbf{X} \in \mathbb{R}^{n \times p}$ be a data matrix of predictor variables and $\mathbf{y} \in \mathbb{R}^n$ be a corresponding vector of responses. For our particular problem, we would like to find features in $\mathbf{X}$ that are correlated with the extreme values of $\mathbf{y}$. (For simplicity, we presume without loss of generality that each of the variables are centered and scaled.) In this paper, we consider the context of a linear data generating model, which will be the focus of the theory presented in section 3 and the empirical investigations of section 4. Here, we assume that the data are generated from a simple linear process: $$\mathbf{y}_i  = \mathbf{X}_i \mathbf{\boldsymbol{\beta}}^* + \epsilon_i, \, \epsilon \text{ i.i.d.}.$$ In order to produce large magnitude extreme values in the observed response $\mathbf{y}_i$ from this model, either some of the corresponding predictors at the observed time $\mathbf{X}_i$ need to be large in magnitude relative or some of the parameters in $\mathbf{\boldsymbol{\beta}}^*$ need to be large in magnitude. 

Note that we assume that the errors $\epsilon_i$ are independently and identically distributed, but do not necessarily assume that they follow a Gaussian distribution. In section 3, we will study both the cases where $\epsilon$ follows a Gaussian distribution and where $\epsilon$ follows a generalized normal, or Subbotin, distribution \citep{sub}. The generalized normal distribution is defined as $$f(\boldsymbol{\epsilon}) = \frac{\gamma}{2\sigma\Gamma(1/\gamma)} e^{-\left(\frac{|\epsilon|}{\sigma}\right)^{\gamma}}$$ for scale parameter $\sigma > 0$ and shape parameter $\gamma> 0$. When $\gamma = 2$, the generalized normal distribution will be equivalent to a Gaussian distribution, while when $\gamma > 2$ the generalized normal distribution will have a thinner tail compared to a Gaussian. Thus, we are specifically interested in studying the case where the generalized normal distribution with $\gamma >2$ as a potential error distribution of the data generating model, as this relatively discourages the presence of extremely large residuals in the regression model estimate when compared to a Gaussian error distribution.

To get estimates of the parameters of the model above, we propose to use the extreme value linear regression model, which is characterized by the $\ell_{\gamma}$-norm regression for $\gamma > 2$. The foundation for this method is a generalized linear model applied to the generalized normal distribution as described above. It follows naturally from the Gaussian case that estimating the parameters of the generalized normal distribution for a particular value of $\gamma$ is analogous to minimizing an $\ell_{\gamma}$ norm regression model loss function \citep{lp1}, which is of the form $$ \mathcal{L}(\mathbf{y}, \mathbf{X}, \hat{\mathbf{\boldsymbol{\theta}}}) =  \frac{1}{\gamma N}\|\mathbf{y} - \mathbf{X} \hat{\mathbf{\boldsymbol{\theta}}} \|_{\gamma}^{\gamma}$$ where $\gamma$ corresponds to the shape parameter in the generalized normal distribution.  As follows from above, we are particularly interested in the case of  $\ell_{\gamma}$ norm regression for $\gamma > 2$.

\begin{figure}[t]
  \centering
  \includegraphics[width=0.9\linewidth]{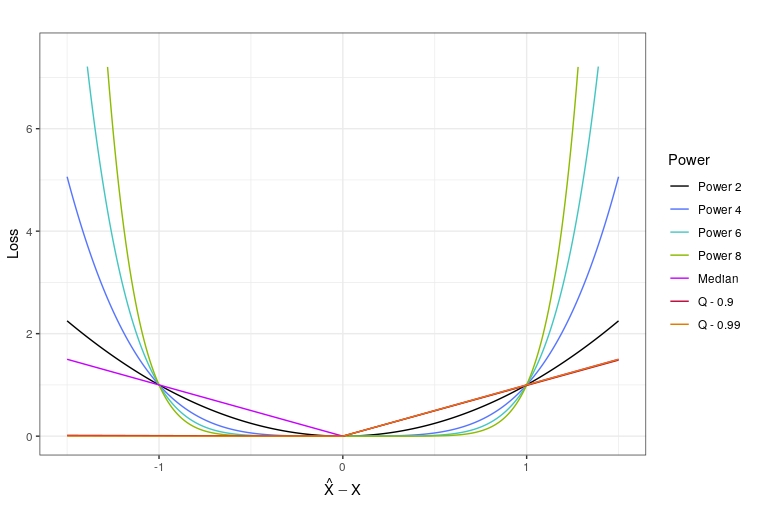}
   \caption{Loss functions for ordinary linear regression, extreme $\ell_{\gamma}$ norm regression, and quantile regression.}
   \label{fig:errplot}
\end{figure}

To demonstrate the differences between the different regression methods discussed in Section 1, we show the respective loss functions for each in Figure \ref{fig:errplot}. Specifically, we show the extreme linear regression loss function for $\gamma = 4, 6, $ and $8$ and the loss for quantile regression at the 0.5 and 0.99 quantiles are shown. Comparing the different methods, we see the advantage that the $\gamma$-th power error loss has over the other two loss functions. Relative to the squared error loss function, the extreme value loss function puts much less weight on very small residuals. However, as the magnitude of a residual increases, the weight given by the extreme value loss function grows exponentially compared to the squared error loss function. In particular, this means that the extreme linear regression will reduce the presence of abnormally large residuals, which occur when there is an extreme value in the response variable which is not captured by the estimate from the model. Thus, the extreme linear regression will find parameter estimates for the model which better predict the occurrences of the extreme values of a response variable. Quantile regression, on the other hand, proportionally increases the relative weight of extreme values by adjusting the weights of an absolute value loss function using linear constants. However, since the loss function only grows linearly, the weight of extreme values compared to relatively large but non-extreme values in the response variable will be small for the quantile regression loss function compared to the extreme linear regression loss function. Additionally, quantile regression can only put increasing weights on either positive or negative residuals in the regression estimate and not both, meaning that it is not suitable in the case where a response variable has both positive and negative extreme values. 

In terms of the impact to the weight of observations in a regression model, our extreme linear regression model functions most similarly to the $\epsilon$-invariant loss used in SVM regression \citep{svm} and to the heterogeneous noise regression models \citep{hetero}. Both of these methods can also be used to substantially decrease the weight of smaller magnitude residuals compared to the larger magnitude ones; this is accomplished by the $\epsilon$-invariant loss by setting the loss for all residuals below a selected magnitude to be 0, while the heterogeneous noise regression models can be used to increase the weight of the observations which are large with respect to either the predictors or the response variable. However, the extreme linear regression model has some notable advantages compared to these techniques; it is not sensitive to the choice of a thresholding hyperparameter as is the case for SVM regression, and it does not require estimation of extra parameters as in the heterogeneous noise model in order to achieve the desired effect for this particular application.

\subsection{Sparse Extreme Value Regression}

In high-dimensional regression problems, automatic feature selection techniques are used to obtain sparse solutions. In many contexts, this is done by adding a sparsity-inducing regularization penalty. In the case of ordinary linear regression, this leads to the penalized squared error loss function. Applying the same idea to the extreme value linear regression model gives the loss function: $$\min_{\mathbf{\boldsymbol{\beta}}} \frac{1}{2N} \|\mathbf{y} - \mathbf{X}\mathbf{\boldsymbol{\beta}}\|_{\gamma}^{\gamma} + \lambda \mathcal{P}(\beta).$$ The form of the extreme value $\ell_{\gamma}$ norm loss function permits the usage of any type of regularization penalty that can be applied to the ordinary linear regression case. For example, one can employ more complex penalties such as SCAD \citep{scad} or MCP \citep{mc}, or specify a more specific structure with penalties such as the Fused Lasso \citep{fs}, Group Lasso \citep{grp}, or Exclusive Lasso \citep{exc}.

{\LinesNumberedHidden
\begin{algorithm}[t]
  \caption{Regularized Extreme Regression Algorithm with Backtracking \label{alg1}}
  \SetAlgoLined
  \SetKwFunction{Union}{Union}\SetKwFunction{FindCompress}{FindCompress}
    \SetKwInOut{Input}{Input}\SetKwInOut{Output}{Output}
  \footnotesize{
  \Input{$\mathbf{y} \in \mathbb{R}^{N \times 1} \mathbf{X} \in \mathbb{R}^{N \times p}$, $\lambda \geq 0, \gamma > 2, \delta > 0, 0 < \alpha < 1.$}
  \BlankLine
  \BlankLine
  \textbf{Initialize:} $\mathbf{\boldsymbol{\beta}}^{(0)} = \boldsymbol{0}_p$\\
  \vspace{5pt}
    \vspace{5pt}
    \While{$\frac{1}{N} \| \boldsymbol{\beta}^{(r)} - \boldsymbol{\beta}^{(r - 1)}\|_1 \geq \delta$}{
    \vspace{5pt}
    \hspace{0.1cm}(1) Find gradient $\nabla g(\mathbf{\boldsymbol{\beta}})$ and optimal step size $t_r$ via backtracking:\\
    \vspace{5pt}
    \hspace{0.4cm}(a) Set $t_r = 1$. \\
    \vspace{5pt}
    \hspace{0.4cm}(b) Calculate $\nabla g^{(r)}(\mathbf{\boldsymbol{\beta}}^{(r)}) = -\gamma \mathbf{X}^T[|y - \mathbf{X}\boldsymbol{\beta}^{(r)}|^{\circ(\gamma - 1)} \circ \text{sgn}(y - \mathbf{X}\boldsymbol{\beta}^{(r)})]$\\
    \vspace{5pt}
    \hspace{0.4cm}(c) \textbf{Repeat:} \\
\vspace{5pt}
\hspace{0.7cm}(i) $\mathbf{z} = \text{prox}_{\lambda * t_r \mathcal{P}}(\boldsymbol{\beta}^{(r)} - t_rg^{(r)}(\mathbf{\boldsymbol{\beta}}^{(r)}))$ \\
\vspace{5pt}
\hspace{0.7cm}(ii) $t_r = \alpha t_r$ \\
\vspace{5pt}
\hspace{0.55cm} \textbf{until} $g(\mathbf{z}) \leq g(\boldsymbol{\beta}^{(r)}) - \nabla g(\boldsymbol{\beta}^{(r)})^T(\boldsymbol{\beta}^{(r)} - \mathbf{z}) + \frac{1}{2t_r}\|\mathbf{z} - \boldsymbol{\beta}^{(r)}\|_2^2$ \\
\vspace{5pt}
    \hspace{0.1cm}(2) Update $\mathbf{\boldsymbol{\beta}}^{(r + 1)} = \mathbf{z}.$\\
\vspace{5pt}
    \hspace{0.1cm}(3) Update $r = r + 1.$\\
    }
    
    \Return{$\hat{\mathbf{\boldsymbol{\beta}}} = \boldsymbol{\beta}^{(r)}$}.}
\end{algorithm}}
    
Similar to the Lasso and other penalized ordinary linear regression models, the objective function for the penalized extreme linear regression can be decomposed in to the sum of two convex functions, the residual norm and the penalty terms. Thus, a proximal gradient descent algorithm can be used to estimate $\hat{\mathbf{\boldsymbol{\beta}}}$. \hide{This type of algorithm is a form of projected gradient descent, in which we iteratively alternate between moving in the negative gradient direction of the differentiable norm loss function and projecting back on to the feasible region defined by the coefficient penalty term via the proximal operator of the penalty function.} Algorithmic convergence properties of proximal gradient descent algorithms for penalized linear regression have been well-studied in recent literature. Notably, it has been shown that the proximal gradient algorithm is guaranteed to converge to a minimum. Additionally, because the $\ell_{\gamma}$ loss function is convex for $\gamma > 2$, if the regularization penalty is also convex, then the algorithm is guaranteed to converge to a global solution \citep{admm}. Algorithm \ref{alg1} gives the general outline of the computational methodology. As a practical consideration, one can choose the $\gamma$ parameter for the regression model either by a priori preference or by using a stability selection criteria, which chooses the $\gamma$ value for the regression model that provides the most consistent estimates of the sparse feature set based on a bootstrapping procedure.

\section{Theoretical Results}

In this section, we present theoretical results for the performance of the sparse extreme value regression method introduced previously. Specifically, we focus our studies on the Extreme Lasso, i.e. the extreme value $\ell_{\gamma}$ norm estimator with an $\ell_1$ regularization penalty. We note that, for the following results, we assume $\gamma$ to be a fixed parameter rather than a parameter to estimate. Our analysis below is separated in to two parts. First, we derive high-dimensional and finite-sample performance guarantees for the Extreme Lasso estimator, showing that it is consistent and variable selection consistent under two different error distributions appropriate for the generalized normal.  We then study our method with respect to the concept of influence functions, a statistic to measure the effect of infinitesimal, pointwise contamination of the covariates and response variable on the resulting regression coefficients. Specifically, we formulate the influence function of the Extreme Lasso regression model and use this to demonstrate that the Extreme Lasso method is more heavily skewed toward selecting features associated with extreme values compared to the ordinary Lasso regression method. Formal proofs for all of the statements in Section 3 can be found in the Appendix.

\subsection{Consistency of the Extreme Lasso}

We now present theoretical results for consistency and model selection consistency of the Extreme Lasso. Our results bear similarity to existing results for the consistency of Lasso-regularized M-estimators; the main difference between the results presented here and those in previous works lies in the distributional assumptions of the errors. Specifically, our contribution lies in deriving concentration bounds for sub-Weibull and sub-Gamma random variables. Consider the linear data generating model:
\begin{align*}
    \mathbf{y}_i  =  \bxi^T  \mathbf{\boldsymbol{\beta}}^* + \epsilon_i, \, \epsilon \text{ i.i.d.}. 
\end{align*} 
The Extreme Lasso regression thus solves the optimization problem:
\begin{align*}
     \minimize_{\beta}  \sum_{i=1}^n  | y_i -  \bxi^T \bbeta |^{\gamma} + \lambda  \| \bbeta \|_1
\end{align*} 
For simplicity, we consider the case when $\gamma$ is an even integer. The problem can now be written as:
\begin{align*}
     \minimize_{\beta}  \sum_{i=1}^n  (y_i -  \bxi^T \bbeta )^\gamma + \lambda  \| \bbeta \|_1
\end{align*} 
Define $$\Ell(\bbeta) = \frac{1}{n} \sum_{i=1}^n \ell (\bxi^T \bbeta - y_i).$$ Clearly, $\Ell$ belongs to the family of M-estimators, whose properties have been widely studied in literature; in particular, \citet{negahban2012unified,loh2017statistical}, and \citet{loh2013regularized} have established the consistency of  M-estimators in the high-dimensional setting. Thus, we can apply the ideas and theories for high-dimensional M-estimators from these papers to the Extreme Lasso case to obtain the results for the regularized extreme value linear regression. 

We first state the previous results regarding the consistency and variable selection consistency for general robust M-estimators which we use below. In the literature, \citet{negahban2012unified} established consistency for high-dimensional M-estimators:

\begin{lemma}[\textbf{Estimation Consistency \citep{negahban2012unified}}] \label{mestconsis}
 Suppose $\Ell$ satisfies the  Restricted Strong Convexity (RSC) condition with curvature $\kappa_{ \Ell}$ and 
\begin{align*}
    \lambda \geq  2 \| \nabla \Ell(\bbeta^*) \|_{\infty} .
\end{align*}
Then $\hat \bbeta$ exists and satisfies the bounds:
\begin{align*}
    \| \hat \bbeta - \bbeta^* \|_2 &\leq \frac{3 \sqrt{s}}{\kappa_{ \Ell}} \lambda
\end{align*} where $s = |\text{supp}(\bbeta^*)|$, i.e., $\|\bbeta^*\|_0$.
\end{lemma} 

\noindent Note that Lemma~\ref{mestconsis} corresponds to Theorem 1 in \citet{negahban2012unified} assuming that the restricted strong convexity (RSC) holds with tolerance parameter $\tau_{ \Ell} = 0$. Also, here we consider $\ell_1$ penalty and $\Psi(\mathcal M) = \sqrt s$. Similarly, \citet{loh2017statistical} established model selection consistency, also known as sparsistency, for high-dimensional robust M-estimators:

\begin{lemma}[\textbf{Model Selection Consistency \citep{lee2015model}}] \label{mestmodelconsis} Suppose the following conditions hold:

(1) $\ell$ satisfies RSC.

(2) $\ell$ satisfies irrepresentability.

\noindent Let $\kappa_{\text{IC}}$  denote the compatibility constant defined in \citet{lee2015model}. 
Then, for any  $\frac{4 \kappa_{\text{IC}}}{\tau} \| \nabla \Ell (\bbeta^*) \|_{\infty} < \lambda < \frac{\kappa_{ \Ell}^2}{2L} \big(2\sqrt s + \frac{\sqrt s}{\kappa_\text{IC}} \frac{\tau}{2} \big)^{-2} \frac{\tau}{\kappa_\text{IC}}$, the optimal solution to an M-estimator problem is unique and model selection consistent: $\hat \beta \in M$. 

Further, if $\min_{a \in \mathcal S} |\beta_a^*| > \frac{2}{\kappa_{ \Ell} } \big( \sqrt{s} + \frac{\tau}{4} \frac{\sqrt s}{\kappa_\text{IC}}\big) \lambda $, then the estimator is also sign consistent: $\text{sign}(\hat \bbeta_{\mathcal S}) = \text{sign} (\bbeta^*_{\mathcal S})$.
\end{lemma}

\noindent Lemma~\ref{mestmodelconsis} refers to Theorem 3.4 in \citet{lee2015model}. The finite constant $\kappa_\text{IC}$ is the compatibility constant between the irrepresentable term and $\rho^*$. $\tau$ is the constant in the irrepresentable condition. Since we consider the $\ell_1$-norm, i.e., $\rho  = \| \cdot \|_1$, we have $k_{\rho} = \sqrt{s}$ and $k_{\rho^*} = 1$ in the theorem. $L$ is a constant such that $\| \nabla^2 \ell(\bbeta) - \nabla^2 \ell(\bbeta^*) \|_2 \leq L \| \bbeta - \bbeta^*\|_2$.  Note in the Lasso problem, it can be shown that $L = 0$; hence there is no upper bound for $\lambda$. In the Extreme Lasso case, in general we have $L \neq 0$ and there is an upper bound for $\lambda$.

Importantly, the results from both Lemma~\ref{mestconsis} and Lemma~\ref{mestmodelconsis} are entirely deterministic. Thus, we can guarantee that, under certain conditions, the extreme value linear regression with the Lasso penalty will provide consistent estimates of the true parameters of the model. Additionally, both Lemma~\ref{mestconsis} and Lemma~\ref{mestmodelconsis} suggest that the key ingredients for statistical consistency are the boundedness of $\| \nabla \Ell(\beta^*)\|_{\infty}$, which ultimately determines the  rate of convergence of $\hat \bbeta$ to $\bbeta^*$ and the local RSC condition. 
Notice that when $\ell$ is the squared error loss, we get the same consistency and model selection consistency rate for the Lasso regression problem:
\begin{align*}
    \| \nabla \Ell(\beta^*)\|_{\infty} = \frac{1}{n} \| \X^T ( y - \X \bbeta^* )\|_{\infty} = \| \X^T \epsilon\|_{\infty}/n.
\end{align*} 
On the other hand, for the Extreme Lasso case, i.e. $ \ell (\bxi^T \bbeta - y_i) =  (y_i - x_i^T \bbeta)^{\gamma}$, we have:
\begin{align*}
    \| \nabla \Ell(\beta^*)\|_{\infty} 
    = \gamma \cdot \frac{1}{n} \| \X^T \epsilon^{\circ (\gamma-1)}      \|_{\infty} .
\end{align*}

To establish complete results for consistency and model selection consistency for the Extreme Lasso, we first build a concentration bound for the quantity $\| \nabla \Ell(\beta^*)\|_{\infty}$, i.e., $\gamma \cdot \frac{1}{n} \| \X^T  \epsilon^{\circ (\gamma-1)}        \|_{\infty}.$ To do this, we first need to build a tail bound for $\epsilon_i^{\gamma-1}$, which will differ under different distributional assumptions on the covariates and error terms in the linear model. These assumptions on the distributional properties will come into play in verifying that the inequality and the RSC condition hold with high probability under the prescribed sample size scaling.  We can then combine the tail bound results with Lemma~\ref{mestconsis} and Lemma~\ref{mestmodelconsis} to derive full results. Below, we present tail bounds for $\frac{1}{n} \| \X^T  \epsilon^{\circ (\gamma-1)}   \|_{\infty} $ under two different distribution assumptions on the error $\epsilon$.  

\subsubsection{Sub-Gaussian Errors}

We first assume that $\epsilon_i$ follows a sub-Gaussian distribution, and we construct a tail bound for a sub-Gaussian random variable raised to a power. 

\begin{lemma}[\textbf{Tail Bound for Sub-Gaussian Raised to a Power}] For sub-Gaussian random variable $\Q$, we have
\begin{align*}
    \mathbb P( |\Q|^{\gamma - 1} \geq t)  \leq 2 \exp\bigg\{- \frac{ t^{2/(\gamma - 1)}}{2\sigma^2} \bigg\}.
\end{align*}\label{subweibulltail}
\end{lemma}

\noindent Under ordinary least squares, i.e. when $\gamma = 2$, we get the usual sub-Gaussian tail bound; when $\gamma = 3$, $\Q^2$ follows a sub-exponential distribution. When $\gamma \geq 4$, as we have for the Extreme Lasso, $\Q^{\gamma - 1}$ is neither sub-Gaussian nor sub-exponential. Instead, in this situation the tail bound will follow what is known in the literature as a sub-Weibull distribution \citep{kuchibhotla2018moving,vladimirova2019sub}, which we define below.

\begin{definition}[\textbf{Sub-Weibull Variables}]
A random variable $\Z$ is said to be sub-Weibull of order $\alpha>0,$ denoted as sub-Weibull($\alpha$), if
$$
\|\Z\|_{\psi_{\alpha}}<\infty, \quad \text { where } \psi_{\alpha}(x):=\exp \left(x^{\alpha}\right)-1 \quad \text { for } x \geq 0.
$$ 
\end{definition} 
\bigskip
\noindent Based on this definition, it follows that if $\Z$ is sub-Weibull $(\alpha),$ then
$$\mathbb{P}(|\Z| \geq t) \leq 2 \exp (-\frac{t^{\alpha}}{\|\Z\|_{\psi_{\alpha}}^{\alpha}}), \text { for all } t \geq 0.$$ In the Extreme Lasso problem, since $\epsilon_i$ is sub-Gaussian, we have $\mathbb P( |\epsilon_i|^{\gamma - 1} \geq t)  \leq 2 \exp\bigg\{- \frac{ t^{2/(\gamma - 1)}}{2\sigma^2} \bigg\}$, which means $\epsilon_i^{\gamma - 1}$ is sub-Weibull, i.e., $ \| \epsilon_i^{\gamma - 1} \|_{\psi_{2/(\gamma - 1)}} < \infty$. In the literature, \citet{kuchibhotla2018moving}  established concentration inequalities related to sub-Weibull random variables. We apply the results and build a tail bound for $\| \sum_{i=1}^n \bxi \epsilon_i^{\gamma-1} \|_{\infty}/n$, i.e., $\| \X^T \epsilon^{\circ (\gamma-1)} \|_{\infty}/n$ by making the substitution $\Z = \epsilon_i^{\gamma - 1}$. Note that by \citet{negahban2012unified}, restricted strong convexity (for M-estimators) with respect to the $\ell_2$-norm is equivalent to the restricted eigenvalues condition (for the Lasso estimator).

\medskip

\begin{lemma}[\textbf{Concentration Bound for Sum of Sub-Weibull Random Variables \citep{kuchibhotla2018moving}}] Consider the Lasso estimator for linear regression case. 
Suppose there exists $0 < \alpha \leq 2$, and $\gamma, K_{n,p} > 0$ such that 
$$ \max \bigg\{ \| X_i \|_{M,\psi_{\alpha}}, \| \epsilon_i \|_{\psi_\gamma} \bigg \} \leq K_{n,p} \hspace{5mm} \text {for all} \hspace{2mm} 1 \leq i \leq n.$$
Also suppose $n \geq 2$, $k \geq 1$ and the covariance matrix $\Sigma_n$ satisfies $\lambda_{\min}(\Sigma_n) \geq K_{n,s}$. Then,  with probability at least $1- 3(np)^{-1} $, 
$$
\left\| \frac{1}{n} \sum_i^n X_i \epsilon_i \right \|_{\infty} \leq 7 \sqrt{2} \sigma_{n, p} \sqrt{\frac{\log (n p)}{n}}+\frac{C_{\tau} K_{n, p}^{2}(\log (2 n))^{1 / \tau}(2 \log (n p))^{1 / \tau}}{n}
$$

where $\frac{1}{\tau} = \frac{1}{\alpha} + \frac{1}{\gamma}$. \label{sumofsubweibulltail}
\end{lemma}

\medskip

\begin{theorem}[\textbf{Consistency for Sub-Gaussian Error}] \label{subgauconsis}
Given the Extreme Lasso program with regularization parameter $\lambda_n  =  2 \gamma \big( 7 \sqrt{2} \sigma_{n, p}   \sqrt{\frac{\log (n p)}{n}}+$ 

\noindent $\frac{C_{\tau} K_{n, p}^{2}(\log (2 n))^{1 / \tau}(2 \log (n p))^{1 / \tau }}{n}\bigg)$, then with probability at least $ 1 - 3 (np)^{-1} $, any optimal solution $\hat \beta$ satisfies the bounds: 
\begin{align*}
    \| \hat \bbeta - \bbeta^* \|_2 &\leq \frac{6 \sqrt{s}}{\kappa_{ \Ell}} \cdot   \gamma \bigg( 7 \sqrt{2} \sigma_{n, p} \sqrt{\frac{\log (n p)}{n}}+\frac{ C_{\tau} K_{n, p}^{2}(\log (2 n))^{1 / \tau}(2 \log (n p))^{1 / \tau }}{n} \bigg).
\end{align*}

where $\tau = 2/(\gamma-1)$.
\end{theorem}


\medskip 

\begin{theorem}[\textbf{Model Selection Consistency for Sub-Gaussian Error}] \label{subgaumodelconsis} Consider the Extreme Lasso program with sub-Gaussian error.  Assume that the loss $\ell$ satisfies Restricted Strong Convexity and covariance matrices satisfy irrepresentability. Consider the family of regularization parameters $\lambda = \frac{4  \kappa_{\text{IC}}}{\tau}  \cdot \gamma \bigg( 7 \sqrt{2} \sigma_{n, p}  \sqrt{\frac{\log (n p)}{n}}+\frac{ C_{\tau} K_{n, p}^{2}(\log (2 n))^{1 / \tau}(2 \log (n p))^{1 / \tau }}{n}\bigg),$ then the following properties holds with probability greater than   $ 1 - 3 (np)^{-1} $: 

(i) The Lasso has a unique solution with support contained
within $S$, i.e. $S(\hat \beta) \subset S(\beta^*)$.

(ii) If $\min_{a \in S} | \beta^{*}_{a}| > (  \frac{\tau}{ \kappa_{\text{IC}}} \cdot \frac{1}{4} + 1 ) \cdot \frac{2\sqrt{s}}{\kappa_{ \Ell}} \cdot \frac{4  \kappa_{\text{IC}}}{\tau}  \cdot \gamma \bigg[ 7 \sqrt{2} \sigma_{n, p}   \sqrt{\frac{\log (n p)}{n}}+\frac{ C_{\tau} K_{n, p}^{2}(\log (2 n))^{1 / \tau}(2 \log (n p))^{1 / \tau }}{n} \bigg] $ , the lasso estimator is also sign consistent: $\text{sign} (\hat \beta_S) = \text{sign} (\beta^{*}_S)$.
\end{theorem} 

\bigskip

Applying the result of Theorem~\ref{subgauconsis} for $\gamma = 2$, we can achieve the usual consistency rate  of $\sqrt{k \log p / n}$ for the ordinary squared error Lasso loss function under the constraint
$$ K_{\varepsilon, r}(\log (n p))^{-1 / 2}(\log (2 n))^{1 / 2}=o\left(n^{1 / 2}\right)
$$ Note that the probability of the bound being satisfied approaches 1 as $n \to \infty$, and thus the bound is proportional $\log(np)$ instead of the usual $\log p$. By setting the probability to be $1 - O(p^{-1})$, the usual Lasso rate  $\sqrt{k \log p / n }$ can be recovered.

\subsubsection{Subbotin Error}

 In the following section, we assume that $\epsilon$ follows a Subbotin distribution, i.e., $\epsilon \sim  \text{Subbotin}(\gamma)$. We study this particular distributional assumption as the Extreme Lasso problem is equivalent to minimizing the negative log-likelihood of the Subbotin distribution plus the regularization penalty. To see this, recall the likelihood of Subbotin distribution:
\begin{align*}
    f_Y(\y;\X;\bbeta) &= c_1     \prod_{i=1}^n \exp \bigg[ -   |y_i - \bxi^T \bbeta |^{\gamma} \bigg]  \\
    &= c_1  \exp \bigg[ - \sum_{i=1}^n |y_i - \bxi^T \bbeta |^{\gamma} \bigg] 
\end{align*}
Thus, the negative log-likelihood, $\ell(\bxi^T \bbeta - y_i) \propto  \sum_{i=1} |y_i - \bxi^T \bbeta |^{\gamma}$, corresponds to the loss function in the Extreme Lasso problem.  Similar to before, our goal is to build a tail bound for  $\| \X^T \epsilon^{\circ (\gamma-1)} \|_{\infty}/n$. To do this, we first observe that $\epsilon_i^{\gamma}$ follows a Gamma distribution.

\begin{lemma}[\textbf{Change of Variables}] \label{changeofvariable}
Suppose $\Z \sim $ Subbotin($\alpha$), where $\alpha$ is an even integer, then $$\Y= \Z^{\alpha} \sim Gamma(\frac{1}{\alpha},1).$$
\end{lemma}

\medskip

\noindent Thus, by Lemma~\ref{changeofvariable}, we have $\epsilon_i^{\gamma} \sim \text{Gamma}(\frac{1}{\theta},1)$. Hence, $\epsilon_i^{\gamma}$ follows a Gamma distribution and can be bounded by sub-Gamma tail bounds in literature \citep{boucheron2013concentration}. These results are stated in Lemma~\ref{subgammatail} and used to derive the results for Theorem~\ref{subbotinconsis} and Theorem~\ref{subbotinmodelconsis} below.

\medskip

\begin{lemma}[\textbf{Concentration Bound for Sub-Gamma Random Variables \citep{boucheron2013concentration}}] \label{subgammatail}
If $\Z \sim $ Gamma($\alpha,\beta$), then we have:
\begin{align*}
    \mathbb P [\Z - \mathbb E \Z] \geq \sqrt{2 \gamma t} + ct ] \leq e^{-t} \hspace{5mm} 
\end{align*}
where $\gamma = \alpha \beta^2$, $c = \beta$. We call that $\Z$ is sub-Gamma with $(\gamma,c)$.
\end{lemma}

\begin{lemma}[\textbf{Concentration Bound for Sum of Sub-Gamma Random Variables}] \label{sumofsubgammatail}
If $\Z \sim $ Gamma($\alpha,\beta$), then  with probability at least $1 - c_1 \exp(- c_2 \log p)$, 
$$
\left\| \frac{1}{n} \sum_i^n X_i \epsilon_i \right \|_{\infty} \leq  \sqrt{ \frac{\log p}{n}} \bigg[    2 \sqrt{ \frac{2}{\gamma}}   +   \sqrt{ \frac{\log p}{n}}   \bigg] 
$$
\end{lemma}

\bigskip

\begin{theorem}[\textbf{Consistency for Subbotin Error}]  \label{subbotinconsis}  Given the Extreme Lasso program with regularization parameter $\lambda_n  =  2 \gamma \sqrt{ \frac{\log p}{n}} \bigg[    2 \sqrt{ \frac{2}{\gamma}}   +   \sqrt{ \frac{\log p}{n}}   \bigg] $, then with probability at least $1 - c_1 \exp(- c_2 \log p)$, any optimal solution $\hat \beta$ satisfies the bounds: 
\begin{align*}
    \| \hat \bbeta - \bbeta^* \|_2 &\leq  \frac{6 \sqrt{s}}{\kappa_{ \Ell}} \gamma  ( \sqrt{ \frac{\log p}{n}} \bigg[    2 \sqrt{ \frac{2}{\gamma}}   +   \sqrt{ \frac{\log p}{n}}   \bigg]  ).
\end{align*}
\end{theorem}

\medskip

\begin{theorem}[\textbf{Model Selection Consistency for Subbotin Error}]  \label{subbotinmodelconsis} Consider the Extreme Lasso program with Subbotin distributed error.  Assume that the loss $\ell$ satisfies Restricted Strong Convexity and the covariance matrices satisfy irrepresentability. Consider the family of regularization parameters $\lambda =  \frac{4 \kappa_{\text{IC}}}{\tau} \gamma \sqrt{ \frac{\log p}{n}} \bigg[    2 \sqrt{ \frac{2}{\gamma}}   +   \sqrt{ \frac{\log p}{n}}   \bigg] $, then the following properties holds with probability greater than   $1 - c_1 \exp(- c_2 \log p)$: 

(i) The Lasso has a unique solution with support contained
within $S$, i.e. $S(\hat \beta) \subset S(\beta^*)$.

(ii) If $\min_{a \in S} | \beta^{*}_{a}| > (\frac{\tau}{\kappa_{\text{IC}}} \cdot \frac{1}{4} + 1 ) \cdot \frac{2\sqrt{s}}{\kappa_{ \Ell}} \cdot \frac{4 \kappa_{\text{IC}}}{\tau} \gamma \sqrt{ \frac{\log p}{n}} \bigg[    2 \sqrt{ \frac{2}{\gamma}}   +   \sqrt{ \frac{\log p}{n}}   \bigg]$, the lasso estimator is also sign consistent: $\text{sign} (\hat \beta_S) = \text{sign} (\beta^{*}_S)$.
\end{theorem}

\medskip

Note that Gaussian distribution is equivalent to the Subbotin distribution when $\theta = 2$. Thus, in the case where $\epsilon_i$ is a Gaussian random variable, we have by Lemma~\ref{changeofvariable} that $\epsilon_i^{2}$ is Gamma($\frac{1}{2},1)$. Hence, $\epsilon_i^2$ is sub-Gamma with $(\frac{1}{2},1)$. Suppose that $\| \X_j \|_{\infty} \leq 1$, we then have $\X_j^T \epsilon$ is a sub-Gamma$(n/2,1)$ random variable. Thus, it follows from Lemma~\ref{subgammatail} that, in this particular case, we have:
\begin{align*}
    \mathbb P\big( \X_j^T \epsilon  - \mathbb E  [\X_j^T \epsilon]  \geq 2 \sqrt{ n t } + t \big) \leq e^{-t}.
\end{align*}
However, if we instead use known Lasso results for $\epsilon$ with sub-Gaussian tail bounds and set $t = \sigma \sqrt{\frac{c \log p }{n}}$, then we have:
\begin{align*}
    \mathbb P\big( |\X_j^T \epsilon  |/n  \geq  { t }  \big) \leq 2 e^{- \frac{nt^2}{2\sigma^2}}.
\end{align*} In effect, the sub-Gamma tail bound has an extra term compared to the sub-Gaussian bound. This can be seen when comparing the result of Theorem~\ref{subbotinconsis} and Theorem~\ref{subbotinmodelconsis} to the Lasso consistency rate derived using sub-Gaussian tail bounds. Specifically, there is an extra factor of $\frac{\log p}{n}$ in the consistency rate result from Theorem~\ref{subbotinconsis} and Theorem~\ref{subbotinmodelconsis} compared to the regular Lasso consistency rate. This is to be expected given that the sub-Gamma is generally a weaker distributional assumption compared to the sub-Gaussian. However, this does show that the bound for Theorem~\ref{subbotinconsis} and Theorem~\ref{subbotinmodelconsis} is not necessarily tight for any particular values of $\theta$.

\subsection{Influence of Extreme Values}

Here, we demonstrate that our Extreme Lasso estimator is better at selecting features associated with extreme values than the regular Lasso estimator. We do this by utilizing the concept of influence functions, which have been previously proposed in the regression literature as a method for analyzing and quantifying the effect of outliers in data on statistical estimators \citep{hampel1968contributions}. However, in previous works, the influence functions have generally been used in order to demonstrate the robustness of a regression estimator to the outlier observations. In our case, we consider the opposite direction, where we show that the Extreme Lasso estimator is more sensitive to the extreme values and hence tends to select features associated with extreme values more. To do this, we show that the value of influence function of the Extreme Lasso is greater than the Lasso, suggesting that our proposed estimator is affected more by extreme values. 

We follow closely the approach by \citet{wang2013robust}. Denote as $\delta_{\mathbf Z}$ the point mass probability distribution at a fixed point $\mathbf{z}=\left(\mathbf{x}_{0}, y_{0}\right)^{T} \in \mathbb{R}^{p+1}$. Given the distribution $F$ of $(\mathbf{x}, y)$ in $\mathbb{R}^{p+1}$ and proportion $\epsilon \in(0,1)$, the mixture distribution of $F$ and $\delta_{\mathbf{Z}}$ is $F_{\epsilon}=(1-\epsilon) F+\epsilon \delta_{\mathbf{Z}}$. Let
\begin{align*}
\boldsymbol{\beta}_{0}^{*} &=\argmin _{\beta}\left[\left\{\int\left( \| y-\mathbf{X}^{T} \boldsymbol{\beta} \|^{\gamma} \right) \mathrm d F\right\}+\sum_{j=1}^{p} p_{\lambda_{ j}}\left(\left|\beta_{j}\right|\right)\right] \\ 
\end{align*} and 
\begin{align*}
\boldsymbol{\beta}_{\epsilon}^{*} &=\argmin _{\beta}\left[\left\{\int\left( \| y-\mathbf{X}^{T} \boldsymbol{\beta} \|^{\gamma}  \right) \mathrm d F_{\epsilon}\right\}+\sum_{j=1}^{p} p_{\lambda_{ j}}\left(\left|\beta_{j}\right|\right)\right].
\end{align*} For the Lasso and Extreme Lasso, $p_{\lambda_{ j}}\left(\left|\beta_{j}\right|\right) = |\beta_{j}|$. For an exponential-type estimator, the influence function at a point $\mathbf{z} \in \mathbb{R}^{p+1}$ is defined as $$\operatorname{IF}\left(\mathbf{z} ; \boldsymbol{\beta}_{0}^{*}\right)=
\lim _{\epsilon \rightarrow 0^{+}}\left(\boldsymbol{\beta}_{\epsilon}^{*}-\boldsymbol{\beta}_{0}^{*}\right) / \epsilon, $$ as long as the limit exists. We use this definition to derive the specific form of the influence function for the Extreme Lasso; the result is shown below in Theorem \ref{influence_fnc}.

\begin{theorem}[\textbf{Influence Function of Extreme Lasso}] \label{influence_fnc}
For the penalized extreme value regression estimators with $\ell_{\gamma}$-norm loss, the $j$th element of the influence function $\operatorname{IF}_{j}\left(\mathbf{z} ; \boldsymbol{\beta}_{0}^{*}\right)$ has the following form:

$$
\operatorname{IF}_{j}\left(\mathbf{z} ; \boldsymbol{\beta}_{0}^{*}\right) \\
= \begin{cases}
0, & \text { if } \beta_{0 j}^{*}=0, \\
-\Gamma_{j}\left\{ - \gamma r_0^{\gamma - 1} x_0  +v_{2}\right\}, & \text{otherwise,}
\end{cases} 
$$
where $\Gamma_{j}$ denotes the $j$th row of $\left\{A\left(\gamma_{0}\right)  -B\right\}^{-1}, r_{0}=y_{0}-$ $\mathbf{x}_{0}^{T} \boldsymbol{\beta}_{0}^{*}$,
$$
\begin{aligned}
v_{2} &=\left\{p_{\lambda_{1}}^{\prime}\left(\left|\beta_{01}^{*}\right|\right) \operatorname{sign}\left(\beta_{01}^{*}\right), \ldots, p_{\lambda_{ d}}^{\prime}\left(\left|\beta_{0d}^{*}\right|\right) \operatorname{sign}\left(\beta_{0d}^{*}\right)\right\}^{T}, \\
B &=\operatorname{diag}\left\{p_{\lambda_{1}}^{\prime \prime}\left(\left|\beta_{01}^{*}\right|\right), \ldots, p_{\lambda_{ d}}^{\prime \prime}\left(\left|\beta_{0d}^{*}\right|\right)\right\},
\end{aligned}
$$
and
$$
A(\gamma)=\int \mathbf{x} \mathbf{x}^{T}  \gamma (\gamma - 1)  \left(y-\mathbf{x}^{T} \boldsymbol{\beta}_0^{*}\right)^{\gamma - 2}  
\times \mathrm d F(\mathbf{x}, y).
$$
\end{theorem}

\medskip

One important implication of this result is that the Extreme Lasso with $\ell_{\gamma}$ regression is more sensitive to features containing extreme values, as formally stated below in Corollary \ref{influence_coro}.

\medskip

\begin{corollary}\label{influence_coro}
The influence function of  the Extreme Lasso with $\gamma > 2$ is greater than the influence function of Lasso.
\end{corollary}

\medskip

\noindent Corollary \ref{influence_coro} can be shown by using a direct comparison with the Lasso influence function, i.e. the case where $\gamma = 2$. Specifically, we have:
\begin{align*}
    \frac{\operatorname{IF}\left((x, y);T_{\text{Extreme}}, F_{\beta_j}\right)}{\operatorname{IF}\left((x, y);T_{\text{Lasso}}, F_{\beta_j}\right)}   &= 
    \frac{\gamma r_0^{\gamma - 1} x_0 -v_{2}}{2 r_0 x_0        -v_{2}} \cdot \frac{A(2) - B_1}{A(\gamma) - B_1} .
\end{align*} Recall that $\beta_0^*$ is the coefficient of fitting the data without extreme values. Hence, if $x_0$ is an influential point, $r_{0}=y_{0}-\mathbf{x}_{0}^{T} \beta_{0}^{*}$ is sufficiently large, which means in this case that $r_0^{\gamma-1} \gg r_0$ for $\gamma > 2$. 
Hence, $ \frac{\operatorname{IF}\left((x, y);T_{\text{Extreme}}, F_{\beta_j}\right)}{\operatorname{IF}\left((x, y);T_{\text{Lasso}}, F_{\beta_j}\right)} > 1$, i.e., the influence function evaluated at $\gamma > 2$ is greater than evaluated at $\gamma = 2$. Thus, the Extreme Lasso will be more likely to select features associated with large magnitude values of $\mathbf{x}$ and $y$ compared to the ordinary Lasso regression.

\section{Empirical Investigations}

Below, we analyze the performance of sparse extreme value linear regression below on two sets of simulations studies and two real-world case studies.

\subsection{Linear Model Simulation Study}

We first study the performance of our method on a simulation study with data generated from the linear model as described in section 2, i.e. $\mathbf{y}_i = \mathbf{X}_i \mathbf{\boldsymbol{\beta}}^*  + \epsilon_i$. We let $\boldsymbol{\epsilon} \overset{iid}{\sim} Gamma(\alpha, \beta)$ (using the rate parameterization) before centering such that $\bar{\epsilon} = 0.$  The predictor matrices $\mathbf{X}$ contain $n = 1000$ observations and $p = 750$ features. The columns of the matrix are generated as AR(1) processes with variance 1 and a cross-correlation with one other column of $\rho = 0.9$. We then add large positive extreme values to the columns at known observation points; these are different for each column. The true parameter vector $\mathbf{\boldsymbol{\beta}}^*$ is set to have 10 randomly selected nonzero entries. Our goal is to recover the full non-zero support of $\mathbf{\boldsymbol{\beta}}^*$ without recovering false positives. We analyze four different varying simulation specifications: 

\begin{enumerate}
    \item The signal to noise ratio of the extreme events relative to baseline noise, which we denote as $\tau$. 
    \item The number of extreme events added to each of the columns of $\mathbf{X}$.
    \item The distribution of the errors $\boldsymbol{\epsilon}$.
    \item The number of dimensions $p$, holding the number of observations and parameter sparsity level constant.
\end{enumerate} We compare the Extreme Lasso regression model, as defined in section 3, with the ordinary Lasso, $\ell_1$ quantile regression, and Lasso regression after preprocessing the data using data-driven thresholding. We fit the Extreme Lasso regression model, as defined in section 3, using $\gamma = 4$ and $\gamma = 6$. For $\ell_1$ quantile regression, we find parameter estimates at the 0.5, 0.9, 0.99, and 0.999 quantiles. Data-driven thresholding is done by using the adaptive CUSUM method \citep{acusum} to identify extreme values in the response variable and removing any data which does not correspond to those observed extreme values. The number of variables for all methods is selected via approximate oracle sparsity tuning. We use 4 replications for each scenario. The results for each of the simulations studies are shown below using average F-1 scores along with the standard deviations across all replications. The full results, which include F-1 scores, true positive rates, and false positive rates for each of the simulations, as well as comparisons with different regularization penalties for the extreme value linear regression and ordinary linear regression models, can be found in the Appendix.


\paragraph{Scenario 1: Magnitudes of Extreme Values in Response} \hfill \break Here, we change the size of the signal to noise ratio, comparing $\tau = 6, 7, 11$, and $15$.  The results are shown in Table \ref{tab:lmm1}. When the signal to noise ratio of the extreme values is not sufficiently large, none of the methods are able to select the correct features. Similarly, if the signal to noise ratio is large enough, all of the methods except quantile regression are able to pick out the correct features. However, we see that there is a fairly large window of $\tau$ values in which the Extreme Lasso is able to find the correct features while ordinary linear regression and thresholding fail.

\begin{table}[H]

\caption{\label{tab:lmm1}Average F-1 scores, changing relative extreme value magnitudes for the linear model.}
\centering
\begin{tabular}{l|l|l|l|l}
\hline
  & $\tau$ =  6 & $\tau$ =  7 & $\tau$ =  11 & $\tau$ =  15\\
\hline
ExLasso ($\gamma = 4$) & 0.196 (0.1382) & 0.209 (0.1778) & 0.875 (0.05) & 0.938 (0.0481)\\
ExLasso ($\gamma = 6$) & 0.296 (0.1416) & 0.782 (0.0894) & 0.85 (0.0577) & 0.938 (0.0481)\\
\hline
Lasso & 0.2 (0.1414) & 0.225 (0.15) & 0.3 (0) & 0.938 (0.0481)\\
\hline
Median & 0.149 (0.0357) & 0.301 (0.2087) & 0.529 (0.0626) & 0.44 (0.1056)\\
Q0.9 & 0.149 (0.0357) & 0.127 (0.0429) & 0.185 (0.1239) & 0.147 (0.0508)\\
Q0.99 & 0.095 (0.0394) & 0.09 (0.0194) & 0.102 (0.0355) & 0.111 (0)\\
Q0.999 & 0.132 (0.1028) & 0.219 (0.2222) & 0.328 (0.2583) & 0.321 (0.1821)\\
\hline
Threshold & 0.028 (0.0556) & 0 (0) & 0 (0) & 0.893 (0.1056)\\
\hline
\end{tabular}
\end{table}

\hide{\begin{figure}[H] 

   \centering
   \includegraphics[scale = 0.5]{img/lm1.jpeg}
   \caption{F-1 scores for changing extreme event magnitudes.}
   \label{fig:lm1}
\end{figure}}


\paragraph{Scenario 2: Number of Extreme Events in Response} \hfill \break We now vary the number of extreme value events $E$ from 1 to 4, with $\tau = 6$. Results are shown in Table \ref{tab:lmm2}. As we observed above, in the case of one extreme event at $\tau = 6$, none of the methods do well. When there is more than one extreme event though, the Extreme Lasso is able to pick out the correct features. None of the other methods are able to perform nearly as well when we increase the number of extreme value events in this case, with only a slight improvement in performance at $E = 4$ compared to  $E = 1$.

\begin{table}[H]

\caption{\label{tab:lmm2}Average F-1 scores, changing number of extreme events for the linear model.}
\centering
\begin{tabular}{l|l|l|l|l}
\hline
  & E =  1 & E =  2 & E =  3 & E =  4\\
\hline
ExLasso ($\gamma = 4$) & 0.875 (0.05) & 0.225 (0.05) & 0.79 (0.0838) & 0.913 (0.0857)\\
ExLasso ($\gamma = 6$) & 0.85 (0.0577) & 0.788 (0.2022) & 0.779 (0.1447) & 0.85 (0.1291)\\
\hline
Lasso & 0.3 (0) & 0.36 (0.1925) & 0.339 (0.0773) & 0.325 (0.05)\\
\hline
Median & 0.529 (0.0626) & 0.513 (0.059) & 0.472 (0.1155) & 0.457 (0.1337)\\
Q0.9 & 0.185 (0.1239) & 0.301 (0.0809) & 0.311 (0.157) & 0.414 (0.1092)\\
Q0.99 & 0.102 (0.0355) & 0.107 (0.0053) & 0.099 (0.0048) & 0.126 (0.0376)\\
Q0.999 & 0.328 (0.2583) & 0.232 (0.0992) & 0.334 (0.0793) & 0.445 (0.2531)\\
\hline
Threshold & 0 (0) & 0 (0) & 0 (0) & 0.075 (0.15)\\
\hline
\end{tabular}
\end{table}

\hide{\begin{figure}[H] 

    \centering
    \includegraphics[scale = 0.5]{img/lm2.jpeg}
    \caption{F-1 scores for different numbers of extreme events.}
    \label{fig:lm2}
\end{figure}}


\paragraph{Scenario 3: Error Distributions} \hfill \break In this scenario, we change the distribution of the added errors by changing the rate parameter of the pre-centered Gamma distribution from which they are generated. By decreasing the rate parameter, we increase the variance of the errors and thus increase the probability of the presence of added errors with magnitudes that are approximately as large as the true extreme events themselves. We study the cases where $\beta = 0.33, 0.2, 0.125$, and $0.083$ at $\tau = 11$. We can see from Table \ref{tab:lmm3} that, starting from the baseline scenario with $\beta = 0.33$, the increasing rate parameter significantly affects the Extreme Lasso in terms of accuracy compared to the other methods. This is not surprising, since we would expect the Extreme Lasso to be more sensitive to large errors that are not actually true signal. However, even in the scenario with the largest error variance, the Extreme Lasso still outperform all of the others. Thus, even in the presence of potentially large residuals, the Extreme Lasso is still a preferable method compared to the others.

\begin{table}[H]

\caption{\label{tab:lmm3}Average F-1 scores, changing error distribution for the linear model.}
\centering
\begin{tabular}{l|l|l|l|l}
\hline
  & $\beta$ =  0.33 & $\beta$ =  0.2 & $\beta$ =  0.125 & $\beta$ =  0.083\\
\hline
ExLasso ($\gamma = 4$) & 0.875 (0.05) & 0.8 (0.1155) & 0.625 (0.1258) & 0.275 (0.2217)\\
ExLasso ($\gamma = 6$) & 0.85 (0.0577) & 0.75 (0.1732) & 0.682 (0.1284) & 0.425 (0.15)\\
\hline
Lasso & 0.3 (0) & 0.2 (0.1414) & 0.262 (0.1103) & 0.175 (0.15)\\
\hline
Median & 0.529 (0.0626) & 0.338 (0.1134) & 0.46 (0.1078) & 0.403 (0.1414)\\
Q-0.9 & 0.185 (0.1239) & 0.122 (0.0465) & 0.121 (0.041) & 0.097 (0.0071)\\
Q-0.99 & 0.102 (0.0355) & 0.092 (0.0055) & 0.092 (0.0096) & 0.097 (0.0096)\\
Q-0.999 & 0.328 (0.2583) & 0.202 (0.1506) & 0.093 (0.0087) & 0.093 (0.0105)\\
\hline
Threshold & 0 (0) & 0.05 (0.1) & 0.073 (0.0994) & 0 (0)\\
\hline
\end{tabular}
\end{table}

\hide{\begin{figure}[H] 

    \centering
    \includegraphics[scale = 0.5]{img/lm3.jpeg}
    \caption{F-1 scores for changing error distributions.}
    \label{fig:lm3}
\end{figure}}


\paragraph{Scenario 4: Number of Dimensions}\hfill \break We change the number of dimensions of the model matrix to study the performance of the different methods in relatively higher dimensional settings. We let $P = 750, 1500, 2250, $ and $3000$, while we hold the number of true features constant (thus decreasing the sparsity level as we increase $P$). Table \ref{tab:lmm4} shows the results. All of the approaches do tend to decay in accuracy. In particular, the least squares and the Extreme Lasso methods tend to show a relatively larger decline in performance, while quantile regression at large quantiles and thresholding appear to be more stable. Once again though, even when performance decays in the higher dimensional settings, the F-1 scores for the Extreme Lasso still exceeds any of the others.

\begin{table}[H]

\caption{\label{tab:lmm4}Average F-1 scores, changing number of dimensions of predictor matrix in the linear model.}
\centering
\begin{tabular}{l|l|l|l|l}
\hline
  & P =  750 & P =  1500 & P =  2250 & P =  3000\\
\hline
ExLasso ($\gamma = 4$) & 0.875 (0.05) & 0.75 (0.0577) & 0.827 (0.0848) & 0.627 (0.2906)\\
ExLasso ($\gamma = 6$) & 0.85 (0.0577) & 0.65 (0.1) & 0.642 (0.1962) & 0.55 (0.1)\\
\hline
Lasso & 0.3 (0) & 0 (0) & 0 (0) & 0 (0)\\
\hline
Median & 0.529 (0.0626) & 0.249 (0.0671) & 0.247 (0.1579) & 0.103 (0.0269)\\
Q-0.9 & 0.185 (0.1239) & 0.117 (0.0553) & 0.102 (0.0119) & 0.103 (0.0269)\\
Q-0.99 & 0.102 (0.0355) & 0.1 (0.0041) & 0.089 (0.0051) & 0.095 (0.0037)\\
Q-0.999 & 0.328 (0.2583) & 0.117 (0.0495) & 0.279 (0.2265) & 0.089 (0.0051)\\
\hline
Threshold & 0 (0) & 0 (0) & 0 (0) & 0 (0)\\
\hline
\end{tabular}
\end{table}

\hide{\begin{figure}[H] 

    \centering
    \includegraphics[scale = 0.5]{img/lm4.jpeg}
    \caption{F-1 scores for changing number of features.}
    \label{fig:lm4}
\end{figure}}

\subsection{Mixture Model Simulation Study}

Next, we study a case where the extreme value linear regression model is misspecified with respect to the data generating model, but where the data still have extreme values. We generate data from a mixture model of the form: $$\mathbf{y}_i = \sum_{k = 1}^K \mathds{1}_{ik}  \mathbf{X}_i \beta_k + \epsilon_i$$ $$\boldsymbol{\epsilon} \overset{iid}{\sim} Gamma(\alpha, \beta).$$ We simulate 4 different sets of predictor variables. The first set contains features which are generated from a mean 0 Gaussian distribution with added extreme values at several randomly selected observation points. The second set contains variables simulated from a Gaussian distribution with no extreme values but which has a mean shift of $2 \sigma^2$ for half of the observations. The third set contains variables which exhibit cross-correlation ($\rho = 0.9$) to one of the variables in the first feature set, but with different extreme value observation points. The fourth set contains uncorrelated white noise variables. We then create a response variable using the above mixture model with $K = 2$ mixture components, where the first component is comprised of the first set of the simulated predictor variables with extreme values, and the second component is comprised of the second set of the simulated predictor variables with mean shift. The first component creates extreme values in the response variable because of their presence in the first set of predictor variables, while the second component will be correlated with the non-extreme values in the response because of the mean shift of the corresponding predictors. Our goal is thus to recover as the support set the variables associated with the first mixture component, i.e. the ones which generate the extreme values in the response, without selecting any variables from any of the others.

The predictor matrices $\mathbf{X}$ we simulate contain $n = 1000$ observations and $p = 750$ columns; 10 features assigned to the each of the first 3 sets of predictor variables as described above and the rest designated as part of the last set. As in the linear regression simulation study, we analyze four different varying simulation specifications:

\begin{enumerate}
    \item The signal to noise ratio of the extreme events relative to baseline noise, $\tau$. 
    \item The number of extreme events added to the variables in the first and third components.
    \item The distribution of the errors $\boldsymbol{\epsilon}$.
    \item The number of dimensions $p$, holding the number of observations and parameter sparsity level for each of the mixture components constant.
\end{enumerate} Again, we compare our method with regularized ordinary least squares regression, $\ell_1$ quantile regression, and Lasso regression after thresholding; we use 4 replications for each scenario, and we compare results using average F-1 scores. Full results can be found in the Appendix.

\paragraph{Scenario 1: Magnitude of Extreme Values of Response Variable} \hfill \break We first vary the size of the signal to noise ratio between $\tau = 6, 7, 9, $ and $50.$ The results are shown in Table \ref{tab:mmm1}. The extreme value methods are able to select the true features at a relatively smaller level of $\tau$. The least squares and thresholding methods are unable to select the features associated with the extreme values until $\tau$ is astronomically large. Meanwhile, the quantile regression methods appear to do better than many of the other methods when $\tau$ is relatively small, but the performance does not improve much with larger values of $\tau$.

\begin{table}[H]

\caption{\label{tab:mmm1}Average F-1 scores, changing relative extreme value magnitudes for the mixture model.}
\centering
\begin{tabular}{l|l|l|l|l}
\hline
  & $\tau$ =  6 & $\tau$ =  7 & $\tau$ =  9 & $\tau$ =  50\\
\hline
ExLasso ($\gamma = 4$) & 0.128 (0.0986) & 0.175 (0.1708) & 0.9 (0.0816) & 1 (0)\\
ExLasso ($\gamma = 6$) & 0.259 (0.3143) & 0.757 (0.0963) & 1 (0) & 1 (0)\\
\hline
Lasso & 0 (0) & 0.075 (0.15) & 0 (0) & 1 (0)\\
\hline
Median & 0.094 (0.0022) & 0.185 (0.1797) & 0.095 (0) & 0.095 (0)\\
Q0.9 & 0.094 (0.0022) & 0.14 (0.0887) & 0.095 (0) & 0.095 (0)\\
Q0.99 & 0.179 (0.0599) & 0.098 (0.0193) & 0.312 (0.1434) & 0.739 (0.1504)\\
Q0.999 & 0.348 (0.0986) & 0.369 (0.1994) & 0.394 (0.1643) & 0.474 (0.1721)\\
\hline
Threshold & 0 (0) & 0.123 (0.1798) & 0.384 (0.392) & 0.977 (0.0455)\\
\hline
\end{tabular}
\end{table}

\hide{\begin{figure}[H] 

    \centering
    \includegraphics[scale = 0.6]{img/mm1.jpeg}
    \caption{F-1 scores for changing extreme event magnitudes.}
    \label{fig:mm1}
\end{figure}}


\paragraph{Scenario 2: Number of Extreme Events in Response} \hfill \break Here, we change the number of extreme value events $E$ from 1 to 4 for $\tau = 6$. Results are shown in Table \ref{tab:mmm2}. As we increase the number of extreme value events, the performance of the extreme value methods steadily increases. Thresholding and quantile regression also tend to perform slightly better with more extreme events, although the improvement is not as drastic. The least squares regression methods never are able to pick any of the features associated with the extreme events. 

\begin{table}[H]

\caption{\label{tab:mmm2}Average F-1 scores, changing number of extreme events for the mixture model.}
\centering
\begin{tabular}{l|l|l|l|l}
\hline
  & E =  1 & E =  2 & E =  3 & E =  4\\
\hline
ExLasso ($\gamma = 4$) & 0.128 (0.0986) & 0.313 (0.1514) & 0.632 (0.1489) & 0.836 (0.0473)\\
ExLasso ($\gamma = 6$) & 0.259 (0.3143) & 0.52 (0.0869) & 0.795 (0.1527) & 0.908 (0.0789)\\
\hline
Lasso & 0 (0) & 0 (0) & 0 (0) & 0 (0)\\
\hline
Median & 0.094 (0.0022) & 0.094 (0.0022) & 0.095 (0) & 0.095 (0)\\
Q0.9 & 0.094 (0.0022) & 0.094 (0.0022) & 0.095 (0) & 0.095 (0)\\
Q0.99 & 0.179 (0.0599) & 0.421 (0.1032) & 0.604 (0.093) & 0.65 (0.1238)\\
Q0.999 & 0.348 (0.0986) & 0.358 (0.0519) & 0.45 (0.1935) & 0.474 (0.1154)\\
\hline
Threshold & 0 (0) & 0.229 (0.1455) & 0.596 (0.1489) & 0.758 (0.1173)\\
\hline
\end{tabular}
\end{table}

\hide{\begin{figure}[H] 

   \centering
    \includegraphics[scale = 0.6]{img/mm2.jpeg}
    \caption{F-1 scores for different numbers of extreme events.}
    \label{fig:mm2}
\end{figure}}


\paragraph{Scenario 3: Error Distributions} \hfill \break In this scenario, we vary the distribution of the added errors by changing the rate parameter to $\beta = 0.33, 0.2, 0.166$, and $0.125$ at $\tau = 9$. Table \ref{tab:mmm3} displays the results. Once again, an increase in the rate parameter significantly degrades the performance the extreme value methods because of the increased presence of large magnitude errors, while other methods are not affected nearly as much. We do eventually see a point where the extreme value methods perform worse than quantile or thresholding. 

\begin{table}[H]

\caption{\label{tab:mmm3}Average F-1 scores, changing error distribution for the mixture model.}
\centering
\begin{tabular}{l|l|l|l|l}
\hline
  & $\beta$  =  0.33 & $\beta$  =  0.2 & $\beta$  =  0.166 & $\beta$  =  0.125\\
\hline
ExLasso ($\gamma = 4$) & 0.9 (0.0816) & 0.875 (0.1258) & 0.816 (0.0526) & 0.278 (0.3587)\\
ExLasso ($\gamma = 6$) & 1 (0) & 1 (0) & 0.582 (0.2852) & 0.184 (0.217)\\
\hline
Lasso & 0 (0) & 0 (0) & 0 (0) & 0 (0)\\
\hline
Median & 0.095 (0) & 0.095 (0) & 0.095 (0) & 0.094 (0.0022)\\
Q-0.9 & 0.095 (0) & 0.095 (0) & 0.095 (0) & 0.094 (0.0022)\\
Q-0.99 & 0.312 (0.1434) & 0.14 (0.0494) & 0.249 (0.0897) & 0.24 (0.0465)\\
Q-0.999 & 0.394 (0.1643) & 0.299 (0.0897) & 0.19 (0.0186) & 0.115 (0.0505)\\
\hline
Threshold & 0.384 (0.392) & 0.05 (0.1) & 0.64 (0.0773) & 0.508 (0.2058)\\
\hline
\end{tabular}
\end{table}

\hide{\begin{figure}[H] 

    \centering
    \includegraphics[scale = 0.6]{img/mm3.jpeg}
    \caption{F-1 scores for changing error distributions.}
    \label{fig:mm3}
\end{figure}}


\paragraph{Scenario 4: Number of Dimensions} \hfill \break We change the number of dimensions of the model matrix to  $P = 750, 1500, 2250, $ and $3000$, with $\tau = 9$ and holding the number of features in components 1, 2, and 3 constant. Results are in Table \ref{tab:mmm4}. The performance of the extreme value and least squares methods do not change much with the increased dimensionality. The quantile regression methods actually perform slightly better with more dimensions, while thresholding tends to do worse.

\begin{table}[H]

\caption{\label{tab:mmm4}Average F-1 scores, changing number of dimensions of predictor matrix in the mixture model.}
\centering
\begin{tabular}{l|l|l|l|l}
\hline
  & P =  750 & P =  1500 & P =  2250 & P =  3000\\
\hline
ExLasso ($\gamma = 4$) & 0.9 (0.0816) & 0.816 (0.0526) & 0.922 (0.0673) & 0.838 (0.0062)\\
ExLasso ($\gamma = 6$) & 1 (0) & 0.947 (0) & 0.961 (0.0263) & 0.947 (0)\\
\hline
Lasso & 0 (0) & 0 (0) & 0 (0) & 0 (0)\\
\hline
Median & 0.095 (0) & 0.095 (0) & 0.095 (0) & 0.095 (0)\\
Q-0.9 & 0.095 (0) & 0.095 (0) & 0.095 (0) & 0.095 (0)\\
Q-0.99 & 0.312 (0.1434) & 0.238 (0.0794) & 0.189 (0.0744) & 0.093 (0.0262)\\
Q-0.999 & 0.394 (0.1643) & 0.39 (0.0962) & 0.501 (0.0663) & 0.577 (0.2264)\\
\hline
Threshold & 0.384 (0.392) & 0.32 (0.1879) & 0.316 (0.087) & 0 (0)\\
\hline
\end{tabular}
\end{table}

\hide{\begin{figure}[H] 

    \centering
    \includegraphics[scale = 0.6]{img/mm4.jpeg}
    \caption{F-1 scores for changing number of features.}
    \label{fig:mm4}
\end{figure}}



\subsection{Real Data Investigation: Calcium Imaging}

We now study the performance of regularized extreme value linear regression on a calcium imaging study from neuroscience, available from the Allen Brain Atlas Brain Observatory data repository \citep{aba}. The data set contains fluorescence traces of neuronal activity for 227 simultaneously recorded neurons in the visual cortex of a mouse brain during periods of controlled visual stimuli. For this study, we work with the parts of the study associated with drifting grating movies, i.e. during time periods which the mice are shown moving black and white gratings of various changing angles and frequencies. Our objective is to predict the recorded fluorescence traces of each of the neurons, with a specific focus on the large positive extreme values that represent neuron firing activity. The predictor variables for the data set are the visual stimulus information from the drifting grating movie, namely the angular orientation and frequency of the drifting gratings being shown to the mouse, as well as other recorded data about the activity of the mouse including treadmill running speed and pupil size and location. We fit the Lasso and 8th power Extreme Lasso regressions to each neuron independently and compare both the chosen stimulus features and the predicted neuron activity traces from each method. Hyperparameter selection for both methods are performed via 5-fold cross validation.

\begin{figure}[t]
    \centering
    \includegraphics[scale = 0.5]{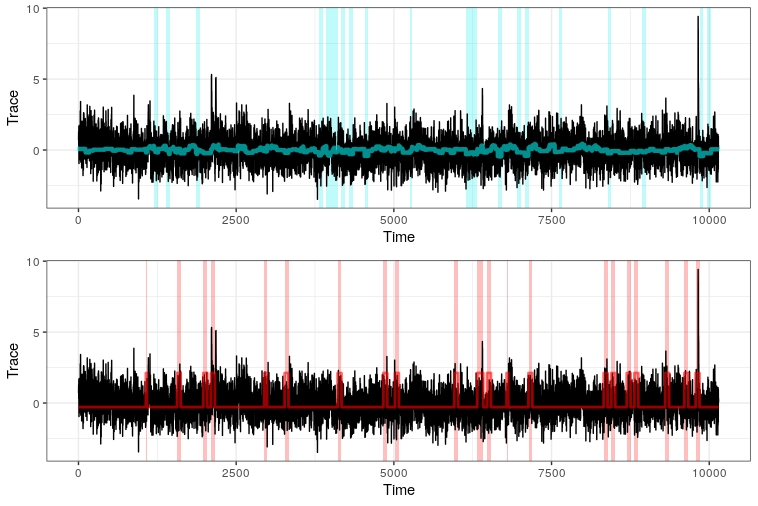}
    \caption{\textbf{Top:} True (black) vs. predicted (blue) neuron trace from the Lasso. Times of chosen angular orientation stimuli are highlighted in color. \textbf{Bottom:} True (black) vs. predicted (red) neuron trace from the Extreme Lasso. Times of chosen angular orientation stimuli are highlighted in color.}
    \label{fig:realdatcal1}
\end{figure}

In Figure \ref{fig:realdatcal1}, we look at the results from one particular neuron. In the top of Figure \ref{fig:realdatcal1}, we see that the prediction from the Lasso does not include any spikes; instead, we see that the Lasso is fitting to the random baseline fluctuations in the fluorescence trace, which are likely to be random noise or measurement artifacts and are not especially useful for this type of data. We also see that the times of the angular orientation stimulus feature selected by the Lasso do not correspond with any spiking activity of the neuron, meaning that the stimuli selected by the Lasso are not particularly scientifically meaningful in this specific context. On the other hand, in the bottom of Figure \ref{fig:realdatcal1}, we see that the Extreme Lasso appears to select an angular orientation stimulus feature for which spikes in the fluorescence trace occur during the time periods where the angle is shown in the drifting grating movie. The predictions from the Extreme Lasso reflect the importance of the extreme values for the estimation procedure as well, with the estimated value of the fluorescence trace being much more sensitive to the spikes in the observed data relative to the Lasso.

\subsection{Real Data Investigation: Climatology}

Our second real data example comes from the field of climatology. The data used here are available from the US EPA AQS Data Mart \citep{epa} and from the MERRA-2 project on NASA MDISC \citep{merra2}. Our goal is to predict and find features associated with large spikes in the hourly measurements of total volatile organic compound (TVOC) concentration in parts per billion (ppb) for a single outdoor monitoring site in Deer Park, Texas. We use as predictor variables for modeling the contemporaneous average hourly data of various atmospheric weather conditions, including temperature, humidity, air pressure, ozone level, wind speed, water vapor concentration, and dew point. From the raw weather data, we also create new predictors using 1 day moving averages of all of the aforementioned variables at time lags ranging from concurrent to 7 days. The data set we look at below contains hourly observations from January 1st, 2015 to December 31st, 2017, totaling approximately 52500 total recorded measurements. We split this in to a training data set which spans the first two years of our data, and a test data set which spans the final year. For this case study, we compare the results from Lasso regression and the 10th power Extreme Lasso regression models. We first perform feature selection with these two methods using the training data set; for this step, hyperparameter values are selected using cross-validation. We then fit the corresponding unbiased regression models to the test data set. Below, we discuss the features which were selected using the regularized regression methods on the training data set, and the model predictions and residuals from the unbiased models on the test data set.

\begin{table}[b]
\begin{tabular}{|c|c|}
\hline
Lasso &  Extreme Lasso \\
\hline
Concurrent hourly air humidity & Concurrent hourly air pressure \\
Concurrent hourly vapor volume & 1 day average precipitation, 5 day lag \\
Concurrent hourly dew point & 1 day average precipitation, 6 day lag \\
Concurrent hourly wind speed & 1 day average precipitation, 7 day lag\\
1 day average temperature, 0 day lag & 1 day average wind speed, 6 day lag \\
1 day average humidity, 0 day lag & \\
1 day average vapor, 0 day lag & \\
\hline
\end{tabular}
\smallskip
\caption{Selected predictors from each regularized regression model.}
\label{tab:featsel}
\end{table}

In Table \ref{tab:featsel}, we show the predictors selected by each of the two methods. As we can see, the Lasso tends to select predictors which are associated with concurrent and current 1-day average atmospheric weather conditions, such as concurrent air humidity and wind speed and 1-day average temperature and water vapor content. On the other hand, the Extreme Lasso mainly selects features associated with daily average weather conditions from 5-7 days prior, particularly with respect to precipitation. Thus, we see that the two methods pick very different sets of predictors. Scientifically, it seems that the Lasso is finding predictors that tend to be associated with smaller common fluctuations in TVOCs, while the Extreme Lasso selects predictors that indicate occurrences of large rainfall events which have been linked in previous literature to large spikes in pollutant concentrations \citep{yvv}.

\begin{figure}[t] 
    \centering
    \begin{minipage}{.5\textwidth}
  \centering
  \includegraphics[width=\linewidth]{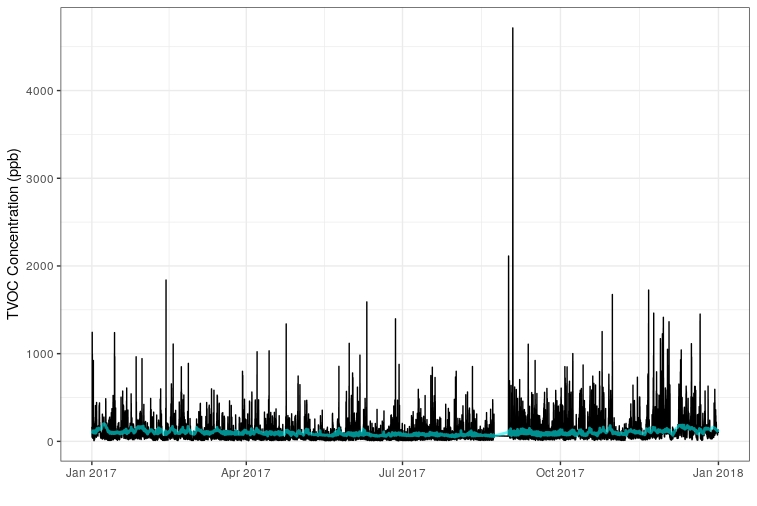}
\end{minipage}%
\begin{minipage}{.5\textwidth}
  \centering
  \includegraphics[width=\linewidth]{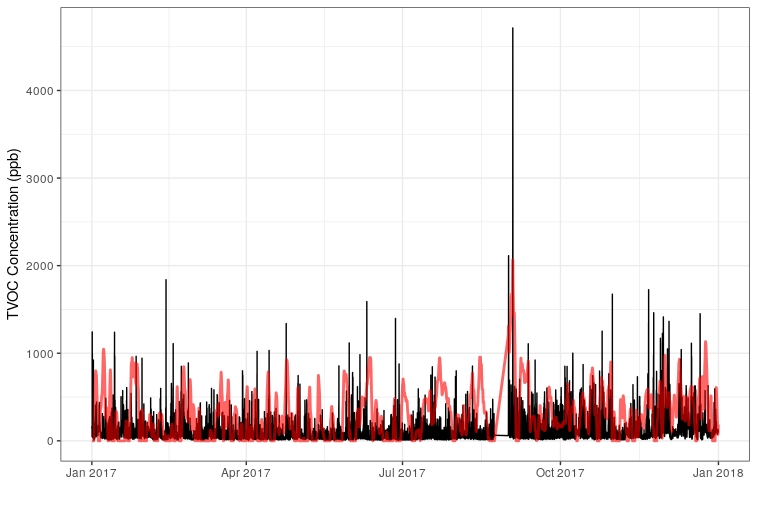}
\end{minipage}
    \begin{minipage}{.5\textwidth}
  \centering
  \includegraphics[width =\linewidth]{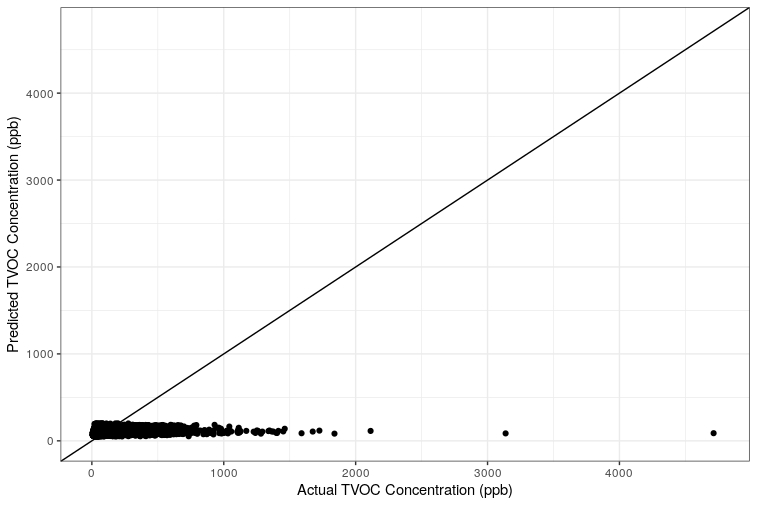}
\end{minipage}%
\begin{minipage}{.5\textwidth}
  \centering
  \includegraphics[width=\linewidth]{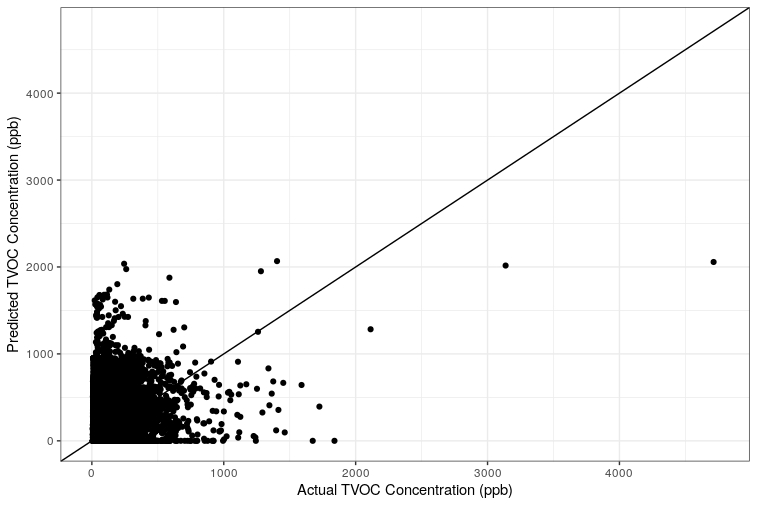}
\end{minipage}
    \caption{\textbf{Top left:} True (black) vs. predicted (blue) hourly TVOC concentration from the ordinary linear regression model. \textbf{Top right:} True (black) vs. predicted (red) hourly TVOC concentration. \textbf{Bottom Left:} True vs. predicted hourly TVOC concentrations from the ordinary linear regression. \textbf{Bottom Right:} True vs. predicted hourly TVOC concentrations from the extreme value linear regression.}
    \label{fig:timepred}
\end{figure}

In Figure \ref{fig:timepred}, we show the predicted TVOC concentrations from the extreme value linear regression and ordinary linear regression models with their previous respective selected features on the test data set; in the top row, we see these plotted over time, and in the bottom we see the predicted and actual values plotted against each other. As we can see, the linear regression model appears to be solely capturing the minor fluctuations which occur regularly across time, but does not seem to capture any of the large spikes in TVOC concentration which occur several times over the course of a year. On the other hand, the extreme value linear regression model, while not always accurate with respect to the smaller value of TVOC, appears to do a much better job in predicting the instances of extreme events where TVOC concentrations spike to irregularly high levels. While neither model is particularly accurate with respect to predicting all of the observed TVOC concentration values, the extreme value linear regression model actually does predict occurrences of extreme value events, whereas the the ordinary linear regression grossly underestimates the TVOC concentration levels when they are above a few hundred parts per billion.

\begin{figure}[t] 
    \centering
  \centering
  \includegraphics[width=\linewidth]{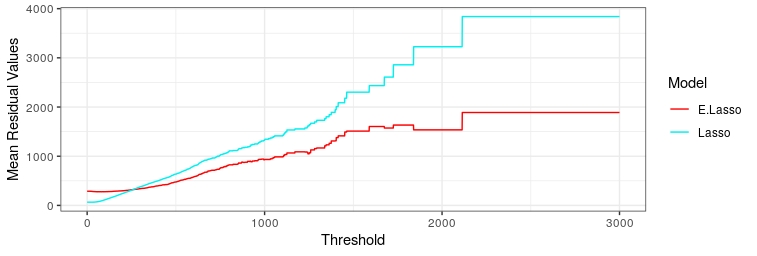}
    \caption{Average absolute value of model residuals from the extreme value linear regression and the ordinary linear regression for values above a concentration threshold.}
    \label{fig:qtplot}
\end{figure}

We analyze more closely the residuals of the extreme of the regression model estimate fit by both methods in Figure \ref{fig:qtplot}. Here, we show the average magnitude of the regression residuals from the ordinary least squares regression and extreme value linear regression models for observations with TVOC concentrations above a changing threshold value. From Figure \ref{fig:qtplot}, we see that the ordinary linear regression predictions are closer to the actual values on average when we consider the entire data set, i.e. when the threshold is 0. However, as we start looking only at data points towards the upper quantiles, we see that the extreme value linear regression begins to outperform the ordinary linear regression in terms of prediction accuracy. For this particular example, the extreme value linear regression model becomes more accurate on average than the ordinary linear regression for observations with TVOC concentration values above 262 ppm, and the difference between the prediction accuracy of the two models gradually increases as we consider smaller, more extreme subsets of the observations of the response variable.

\section{Discussion}

In this paper, we have introduced the extreme value linear regression model, a potential new methodological approach to linear regression for extreme values. Our method is motivated by $\ell_{\gamma}$-norm regression, which gives much more weight to the loss for large magnitude residuals relative to ordinary least squares. This concept has several advantages over other methods currently used in the literature, namely that it does not require using a two-step pipeline of pre-processing the data before analysis, nor does it force the data to be binarized as either extreme or non-extreme. Our method also does not necessitate the a priori choice of certain model hyperparameters that may be difficult to select. Our simulation studies provide promising results which demonstrate that, for a response variable with rare extreme values, the extreme value linear regression model with automatic feature selection performs better than quantile regression, thresholding, and least squares penalized regression in terms of selecting predictors which are correlated with the extreme values in the response.  We have also shown deterministic finite sample performance guarantees for consistency and model selection consistency of the Extreme Lasso regression model under the assumption of a linear data generating model with different potential error distributions, demonstrating that the estimates from the extreme value linear regression model are reliable. The theoretical results here could also be of use for other types of similar problems. In particular, the concentration bounds and theory presented for the case of generalized normal distributed errors for $\gamma > 2$ could be applied to generate new theoretical results for other mathematical statistics  problems. 

There are several potential areas for future work for the extreme value linear regression. Our theoretical work has mainly focused on using a simple $\ell_1$ Lasso penalty for regularization under the linear regression data generating model; however, the extreme values in a response variable could come from a variety of different data generating models. Theoretical results for the variance of estimators in the low-dimensional case have also not been addressed here, and could be of interest for future study. There remains potential methodological developments for the extreme value linear regression to explore as well. Just as ordinary regression methods are insufficient for fitting a model for the extreme values, traditional model selection methods may not work particularly well in this context. While we use regular cross-validation to select $\lambda$ during model fitting in our real data examples, we recognize that this may not be the optimal method. The model selection problem may instead require more nuanced treatment, as naive cross-validation methods may not work well when the extreme values are particularly rare. Though we have described a general approach for selecting $\gamma$ for an individual data set, additional empirical investigation may be useful for gaining a better understanding of what values of $\gamma$ are typically useful for analyzing real-world data. Also, while we have presented a couple of potential applications, our method has the potential to be applied broadly to a variety of fields, such as for signal processing or for spectral domain data; explorations in to other applications could provide new insights in these areas. In conclusion, we develop a novel method for extreme value regression modeling that opens many area for future research.


\appendix

\section{Proofs for Section 3}

\subsection{Lemma 3.3}

For $t > 0$,
\begin{align*}
    \mathbb P(\Q^{\gamma} \geq t)  = \mathbb P(\Q \geq t^{^{1/\gamma}})  = \mathbb P(  e^{\lambda \Q} \geq e^{\lambda t^{^{1/\gamma}}}  ) \leq \frac{e ^{ \sigma^2 \lambda^2 / 2 } }{ e^{\lambda t^{^{1/\gamma}}}}  = \exp \bigg\{ \sigma^2 \lambda^2 / 2  - \lambda t^{^{1/\gamma}} \bigg\}.
\end{align*}
The right hand side is minimized by $\lambda^* = \frac{t^{1/\gamma}}{\sigma^2}$. Hence, we have
\begin{align*}
    \mathbb P(\Q^{\gamma} \geq t)  \leq \exp\bigg\{- \frac{ t^{2/\gamma}}{2\sigma^2} \bigg\}.
\end{align*} \hfill $\square$

\subsection{Theorem 3.1}

In the Extreme Lasso problem, by Lemma~\ref{subweibulltail}, $ \| \epsilon_i^{\gamma - 1} \|_{\psi_\gamma} \leq K_{n,p}$ where $\gamma = \frac{2}{\gamma -1}$. For fixed design $\X$, $\X_i$'s are marginally sub-Weibull $(\infty)$ and
$$
\max _{1 \leq i \leq n}\left\|X_{i}\right\|_{M, \psi_{2}} \leq \max _{1 \leq i \leq n}\left\|X_{i}\right\|_{M, \psi_{\infty}}=\max _{1 \leq i \leq n} \max _{1 \leq j \leq p}\left|X_{i}(j)\right|.
$$ Applying Lemma~\ref{sumofsubweibulltail} with $\alpha=\infty$, we have  $\tau = 2/(\gamma-1)$.	Therefore, by choosing $\lambda_n$ to be 
$$
\lambda_n = 2 \gamma \big( 7 \sqrt{2} \sigma_{n, p}   \sqrt{\frac{\log (n p)}{n}}+ \frac{C_{\tau} K_{n, p}^{2}(\log (2 n))^{1 / \tau}(2 \log (n p))^{1 / \tau }}{n}\bigg),
$$
the Extreme Lasso estimator satisfies 
\begin{align*}
    \| \hat \bbeta - \bbeta^* \|_2 &\leq \frac{6 \sqrt{s}}{\kappa_{ \Ell}} \cdot   \gamma \bigg( 7 \sqrt{2} \sigma_{n, p} \sqrt{\frac{\log (n p)}{n}}+\frac{ C_{\tau} K_{n, p}^{2}(\log (2 n))^{1 / \tau}(2 \log (n p))^{1 / \tau }}{n} \bigg)
\end{align*} where $\tau = 2/(\gamma-1).$ 

\hfill $\square$

\subsection{Theorem 3.2}

Similar to Theorem~\ref{subgauconsis}, we prove model selection Consistency holds by applying Lemma~\ref{mestmodelconsis} with the concentration bound demonstrated in Lemma~\ref{sumofsubweibulltail}.

\hfill $\square$

\subsection{Lemma 3.5}

 Suppose $\Z \sim $ Subbotin($\alpha$), i.e.
\begin{align*}
    f_\Z(z)  = \frac{\alpha}{2 \Gamma(\frac{1}{\alpha} )} \exp\big[ - | z |^{\alpha} \big] .
\end{align*} Let $\Y  =   \Z ^{\alpha} $, then $z = \pm y^{\frac{1}{\alpha}}$, $| \frac{dz}{dy} | = \frac{1}{\alpha} y^{\frac{1}{\alpha} - 1} $ and
\begin{align*}
    f_\Y(y) = \frac{\alpha}{\Gamma(\frac{1}{\alpha})} \exp[-y] \frac{1}{\alpha} y^{\frac{1}{\alpha} - 1} =   \frac{1}{\Gamma(\frac{1}{\alpha})}  \exp[-y]  y^{\frac{1}{\alpha} - 1}.
\end{align*}
Thus, $\Y= \Z^{\alpha} \sim$ Gamma$(\frac{1}{\alpha},1)$.

\hfill $\square$

\subsection{Lemma 3.7}

By Lemma~\ref{changeofvariable}, we have $\epsilon_i^{\theta} \sim$ Gamma$(1/\theta,1)$. Hence, Lemma~\ref{subgammatail} suggests
\begin{align*}
    \mathbb P\big( \epsilon_i^{\theta}  - \mathbb E  [\epsilon_i^{\theta}]  \geq 2 \sqrt{2 \frac{1}{\theta} t } + t \big) \leq e^{-t}.
\end{align*} If $\gamma - 1 \leq \theta$, we can show that $\epsilon_i^{\gamma-1}$ is also sub-Gamma with $(\frac{1}{\theta},1)$ as the latter one has lower tail. 
\begin{align*}
    \mathbb P\big(  \epsilon_i^{\gamma-1}  - \frac{1}{\theta} \geq  2 \sqrt{2 \frac{1}{\theta} t } + t \big)   \leq \mathbb P\big(  \epsilon_i^{\theta}  - \frac{1}{\theta} \geq  2 \sqrt{2 \frac{1}{\theta} t } + t \big)   \leq e^{-t}.
\end{align*} For $\| \X_j \|_{\infty} \leq 1$, we have $\X_j^T \epsilon^{\gamma-1}$ is  sub-Gamma with $(n/\theta,1)$ since sum of sub-Gamma is also sub-Gamma. From this, we find
\begin{align*}
    \mathbb P\big( \X_j^T \epsilon^{\gamma-1}  - \mathbb E  [\X_j^T \epsilon^{\gamma-1}]  \geq 2 \sqrt{2 \frac{n}{\theta} t } + t \big) \leq e^{-t}.
\end{align*}
By using union bounds, we thus have
\begin{align*}
    \mathbb P\big( \| \X^T \epsilon^{\gamma-1}  \|_{\infty} \geq  2 \sqrt{2 \frac{n}{\theta} t } + t \big) \leq p e^{-t} .
\end{align*} Choosing $t = \log p$, we get
\begin{align*}
    \|\X^T \epsilon^{\gamma-1} \|_{\infty} \leq  2 \sqrt{ \frac{2}{\theta}} \sqrt{n \log p} + \log p    
\end{align*}
with probability at least $1 - c_1 \exp(- c_2 \log p)$. This is equivalent to:
\begin{align*}
    \|\X^T \epsilon^{\gamma-1} \|_{\infty}/n  & \leq 2 \sqrt{ \frac{2}{\theta}}  \sqrt{ \frac{\log p}{n}} + \frac{\log p}{n} \\
    & = \sqrt{ \frac{\log p}{n}} \bigg[    2 \sqrt{ \frac{2}{\theta}}   +   \sqrt{ \frac{\log p}{n}}   \bigg] \\
    & \leq  \sqrt{ \frac{\log p}{n}} \bigg[    2 \sqrt{ \frac{2}{\gamma}}   +   \sqrt{ \frac{\log p}{n}}   \bigg] \\
\end{align*} with probability at least $1 - c_1 \exp(- c_2 \log p)$. 

\hfill $\square$

\subsection{Theorem 3.3}

 By applying Lemma~\ref{mestconsis} with the concentration bound demonstrated in Lemma~\ref{sumofsubgammatail}, we have the consistency result. 
 
 \hfill $\square$

\subsection{Theorem 3.4}

 By applying Lemma~\ref{mestmodelconsis} with the concentration bound demonstrated in Lemma~\ref{sumofsubgammatail}, we have the model consistency result.
 
 \hfill $\square$

\subsection{Theorem 3.5}

By the KKT condition as required for optimality of $\boldsymbol{\beta}_{\epsilon}^{*}$, we have:
\begin{align} \label{eq:influ1}
(1-\epsilon) \int\left[ - \gamma \left(y-\mathbf{x}^{T} \boldsymbol{\beta}_{\epsilon}^{*}\right)^{\gamma-1}  \mathbf{x} \right]  \times \mathrm d F(\mathbf{x}, y) \nonumber \\
+\epsilon \left( - \gamma \left(y_0-\mathbf{x}_0^{T} \boldsymbol{\beta}_{\epsilon}^{*}\right)^{\gamma-1}  \times \mathbf{x}_{0} \right)-v_{1}(\epsilon)=0 ,  
\end{align}
where $v_1(\epsilon)=\left(p_{\lambda_{1}}^{\prime}\left(\left|\beta_{\epsilon 1}\right|\right) \operatorname{sign}\left(\beta_{\epsilon 1}\right), \ldots, p_{\lambda_{ d}}^{\prime}\left(\left|\beta_{\epsilon d}\right|\right) \operatorname{sign}\left(\beta_{\epsilon d}\right)\right)^{T}$.  Let $r_{0}=y_{0}-\mathbf{x}_{0}^{T} \beta_{0}^{*}$. Differentiating with respect to $\epsilon$ in both sides of~\eqref{eq:influ1} and letting $\epsilon \rightarrow 0$, we obtain
\begin{align} \label{eq:influ2}
\int\left[   - \gamma (\gamma - 1)   \left  (y-\mathbf{x}^{T} \boldsymbol{\beta}_{\epsilon}^{*}\right)^{\gamma - 2}  \times \frac{\partial}{\partial \epsilon}\left(y-\mathbf{x}^{T} \boldsymbol{\beta}_{\epsilon}^{*}\right) \mathbf{x}         \right]&\left. \mathrm d F(\mathbf{x}, y)\right|_{\epsilon=0}-\frac{\partial v_{1}(\epsilon)}{\partial \epsilon}  \nonumber \\
&= \gamma r_0^{\gamma - 1} x_0                   -v_{2}  ,
\end{align}
where $v_2=\left(p_{\lambda_{1}}^{\prime}\left(\left|\beta_{01}^{*}\right|\right) \operatorname{sign}\left(\beta_{01}^{*}\right), \ldots, p_{\lambda_{ d}}^{\prime}\left(\left|\beta_{ 0d}^{*}\right|\right) \operatorname{sign}\left(\beta_{ 0d}^{*}\right)\right)^{T}$. Using~\eqref{eq:influ1} and~\eqref{eq:influ2}, it can be shown that
$$
\left( A(\gamma) -B_{1}\right)\left[\operatorname{IF}\left\{\left(\mathbf{x}_{0}, y_{0}\right), \boldsymbol{\beta}_{0}^{*}\right\}\right]=\gamma r_0^{\gamma - 1} x_0        -v_{2},
$$
where
$$
A(\gamma)=\int \mathbf{x} \mathbf{x}^{T} \gamma (\gamma - 1)  \left(y-\mathbf{x}^{T} \boldsymbol{\beta}_0^{*}\right)^{\gamma - 2} \times \mathrm d F(\mathbf{x}, y),$$
$$
\begin{aligned}
B_{1}=& \operatorname{diag}\left\{p_{\lambda_{1}}^{\prime \prime}\left(\left|\beta_{01}^{*}\right|\right)+p_{\lambda_{1}}^{\prime}\left(\left|\beta_{01}^{*}\right|\right) \delta\left(\beta_{01}^{*}\right), \ldots, p_{\lambda_{ d}}^{\prime \prime}\left(\left|\beta_{ 0d}^{*}\right|\right) +p_{\lambda_{d}}^{\prime}\left(\left|\beta_{ 0d}^{*}\right|\right) \delta\left(\beta_{0d}^{*}\right)\right\},
\end{aligned}
$$
with
$$
\delta(x)= \begin{cases}
+\infty, & \text { if } x=0, \\
0, & \text { otherwise. }
\end{cases} 
$$

\hfill $\square$

\section{Full Tabular Results}

For our full results, we also compare different regularization penalties for the extreme value linear regression model and the linear regression model.

\subsection{Linear Model Simulation Study}

\noindent{\textit{Scenario 1: Changing Magnitude of Extreme Values of Response Variable}}

\begin{table}[H]

\caption{Average F-1 score for changing extreme value magnitude.}
\centering
\begin{tabular}{l|l|l|l|l}
\hline
  & $\tau$ =  6 & $\tau$ =  7 & $\tau$ =  11 & $\tau$ =  15\\
\hline
ExLasso ($\gamma = 4$) & 0.196 (0.1382) & 0.209 (0.1778) & 0.875 (0.05) & 0.938 (0.0481)\\
ExLasso ($\gamma = 6$) & 0.296 (0.1416) & 0.782 (0.0894) & 0.85 (0.0577) & 0.938 (0.0481)\\
\hline
ExSCAD 4th & 0.196 (0.1382) & 0.209 (0.1778) & 0.875 (0.05) & 0.938 (0.0481)\\
ExSCAD 6th & 0.296 (0.1416) & 0.782 (0.0894) & 0.85 (0.0577) & 0.938 (0.0481)\\
\hline
ExMCP 4th & 0.1 (0.1155) & 0.195 (0.1556) & 0.888 (0.0637) & 0.938 (0.0481)\\
ExMCP 6th & 0.255 (0.0662) & 0.757 (0.1606) & 0.864 (0.0474) & 0.938 (0.0481)\\
\hline
Lasso & 0.2 (0.1414) & 0.225 (0.15) & 0.3 (0) & 0.938 (0.0481)\\
SCAD & 0.2 (0.1414) & 0.225 (0.15) & 0.3 (0) & 0.938 (0.0481)\\
MCP & 0.2 (0.1414) & 0.225 (0.15) & 0.3 (0) & 0.963 (0.0477)\\
\hline
Median & 0.149 (0.0357) & 0.301 (0.2087) & 0.529 (0.0626) & 0.44 (0.1056)\\
Q0.9 & 0.149 (0.0357) & 0.127 (0.0429) & 0.185 (0.1239) & 0.147 (0.0508)\\
Q0.99 & 0.095 (0.0394) & 0.09 (0.0194) & 0.102 (0.0355) & 0.111 (0)\\
Q0.999 & 0.132 (0.1028) & 0.219 (0.2222) & 0.328 (0.2583) & 0.321 (0.1821)\\
\hline
Threshold & 0.028 (0.0556) & 0 (0) & 0 (0) & 0.893 (0.1056)\\
\hline
\end{tabular}
\end{table}

\begin{table}[H]

\caption{Average true positive rates for changing extreme value magnitude.}
\centering
\begin{tabular}{l|l|l|l|l}
\hline
  & $\tau$ =  6 & $\tau$ =  7 & $\tau$ =  11 & $\tau$ =  15\\
\hline
ExLasso ($\gamma = 4$) & 0.193 (0.1355) & 0.196 (0.1571) & 0.875 (0.05) & 0.927 (0.0487)\\
ExLasso ($\gamma = 6$) & 0.293 (0.1421) & 0.767 (0.1054) & 0.85 (0.0577) & 0.927 (0.0487)\\
\hline
ExSCAD 4th & 0.193 (0.1355) & 0.196 (0.1571) & 0.875 (0.05) & 0.927 (0.0487)\\
ExSCAD 6th & 0.293 (0.1421) & 0.767 (0.1054) & 0.85 (0.0577) & 0.927 (0.0487)\\
\hline
ExMCP 4th & 0.1 (0.1155) & 0.191 (0.1488) & 0.877 (0.0517) & 0.927 (0.0487)\\
ExMCP 6th & 0.262 (0.0828) & 0.764 (0.1467) & 0.855 (0.053) & 0.927 (0.0487)\\
\hline
Lasso & 0.2 (0.1414) & 0.225 (0.15) & 0.3 (0) & 0.927 (0.0487)\\
SCAD & 0.2 (0.1414) & 0.225 (0.15) & 0.3 (0) & 0.927 (0.0487)\\
MCP & 0.2 (0.1414) & 0.225 (0.15) & 0.3 (0) & 0.952 (0.0552)\\
\hline
Median & 0.398 (0.2045) & 0.318 (0.16) & 0.446 (0.041) & 0.435 (0.0842)\\
Q0.9 & 0.398 (0.2045) & 0.164 (0.1179) & 0.158 (0.1061) & 0.145 (0.0449)\\
Q0.99 & 0.128 (0.1367) & 0.086 (0.0384) & 0.09 (0.0362) & 0.125 (0)\\
Q0.999 & 0.175 (0.2165) & 0.196 (0.2072) & 0.303 (0.2673) & 0.352 (0.2432)\\
\hline
Threshold & 0.031 (0.0625) & 0 (0) & 0 (0) & 0.864 (0.1174)\\
\hline
\end{tabular}
\end{table}

\begin{table}[H]

\caption{Average false positive rates for changing extreme value magnitude.}
\centering
\begin{tabular}{l|l|l|l|l}
\hline
  & $\tau$ =  6 & $\tau$ =  7 & $\tau$ =  11 & $\tau$ =  15\\
\hline
ExLasso ($\gamma = 4$) & 0.011 (0.0017) & 0.011 (0.0014) & 0.002 (7e-04) & 0.001 (7e-04)\\
ExLasso ($\gamma = 6$) & 0.01 (0.002) & 0.003 (0.0017) & 0.002 (8e-04) & 0.001 (7e-04)\\
\hline
ExSCAD 4th & 0.011 (0.0017) & 0.011 (0.0014) & 0.002 (7e-04) & 0.001 (7e-04)\\
ExSCAD 6th & 0.01 (0.002) & 0.003 (0.0017) & 0.002 (8e-04) & 0.001 (7e-04)\\
\hline
ExMCP 4th & 0.011 (0.0014) & 0.01 (7e-04) & 0.002 (7e-04) & 0.001 (7e-04)\\
ExMCP 6th & 0.01 (0.002) & 0.003 (0.0017) & 0.002 (8e-04) & 0.001 (7e-04)\\
\hline
Lasso & 0.011 (0.0019) & 0.01 (0.002) & 0.009 (0) & 0.001 (7e-04)\\
SCAD & 0.011 (0.0019) & 0.01 (0.002) & 0.009 (0) & 0.001 (7e-04)\\
MCP & 0.01 (8e-04) & 0.01 (0.002) & 0.009 (0) & 0.001 (8e-04)\\
\hline
Median & 0.004 (0.0061) & 0.009 (0.0046) & 0.011 (0) & 0.008 (0.0013)\\
Q0.9 & 0.008 (0.0036) & 0.011 (0.0045) & 0.01 (0.003) & 0.008 (0.0016)\\
Q0.99 & 0.017 (0.0093) & 0.016 (0.0058) & 0.018 (0.0059) & 0.009 (0)\\
Q0.999 & 0.015 (0.0083) & 0.016 (0.0055) & 0.013 (0.0078) & 0.008 (0.0039)\\
\hline
Threshold & 0.009 (0.0019) & 0.011 (0.0017) & 0.01 (7e-04) & 0.002 (0.0017)\\
\hline
\end{tabular}
\end{table}

\newpage

\noindent{\textit{Scenario 2: Changing Number of Extreme Events in Response}}

\begin{table}[H]

\caption{Average F-1 scores for changing number of extreme events.}
\centering
\begin{tabular}{l|l|l|l|l}
\hline
  & E =  1 & E =  2 & E =  3 & E =  4\\
\hline
ExLasso ($\gamma = 4$) & 0.875 (0.05) & 0.225 (0.05) & 0.79 (0.0838) & 0.913 (0.0857)\\
ExLasso ($\gamma = 6$) & 0.85 (0.0577) & 0.788 (0.2022) & 0.779 (0.1447) & 0.85 (0.1291)\\
\hline
ExSCAD 4th & 0.875 (0.05) & 0.225 (0.05) & 0.79 (0.0838) & 0.913 (0.0857)\\
ExSCAD 6th & 0.85 (0.0577) & 0.788 (0.2022) & 0.779 (0.1447) & 0.85 (0.1291)\\
\hline
ExMCP 4th & 0.888 (0.0637) & 0.25 (0.1) & 0.788 (0.1192) & 0.913 (0.0857)\\
ExMCP 6th & 0.864 (0.0474) & 0.813 (0.1555) & 0.779 (0.1447) & 0.89 (0.0978)\\
\hline
Lasso & 0.3 (0) & 0.36 (0.1925) & 0.339 (0.0773) & 0.325 (0.05)\\
SCAD & 0.3 (0) & 0.36 (0.1925) & 0.339 (0.0773) & 0.325 (0.05)\\
MCP & 0.3 (0) & 0.295 (0.0741) & 0.339 (0.0773) & 0.325 (0.05)\\
\hline
Median & 0.529 (0.0626) & 0.513 (0.059) & 0.472 (0.1155) & 0.457 (0.1337)\\
Q0.9 & 0.185 (0.1239) & 0.301 (0.0809) & 0.311 (0.157) & 0.414 (0.1092)\\
Q0.99 & 0.102 (0.0355) & 0.107 (0.0053) & 0.099 (0.0048) & 0.126 (0.0376)\\
Q0.999 & 0.328 (0.2583) & 0.232 (0.0992) & 0.334 (0.0793) & 0.445 (0.2531)\\
\hline
Threshold & 0 (0) & 0 (0) & 0 (0) & 0.075 (0.15)\\
\hline
\end{tabular}
\end{table}

\begin{table}[H]

\caption{Average true positive rates for changing number of extreme events.}
\centering
\begin{tabular}{l|l|l|l|l}
\hline
  & E =  1 & E =  2 & E =  3 & E =  4\\
\hline
ExLasso ($\gamma = 4$) & 0.875 (0.05) & 0.225 (0.05) & 0.782 (0.0894) & 0.902 (0.0818)\\
ExLasso ($\gamma = 6$) & 0.85 (0.0577) & 0.777 (0.1913) & 0.759 (0.1297) & 0.85 (0.1291)\\
\hline
ExSCAD 4th & 0.875 (0.05) & 0.225 (0.05) & 0.782 (0.0894) & 0.902 (0.0818)\\
ExSCAD 6th & 0.85 (0.0577) & 0.777 (0.1913) & 0.759 (0.1297) & 0.85 (0.1291)\\
\hline
ExMCP 4th & 0.877 (0.0517) & 0.25 (0.1) & 0.777 (0.0997) & 0.902 (0.0818)\\
ExMCP 6th & 0.855 (0.053) & 0.802 (0.1436) & 0.759 (0.1297) & 0.882 (0.1133)\\
\hline
Lasso & 0.3 (0) & 0.333 (0.1414) & 0.329 (0.0583) & 0.325 (0.05)\\
SCAD & 0.3 (0) & 0.333 (0.1414) & 0.329 (0.0583) & 0.325 (0.05)\\
MCP & 0.3 (0) & 0.291 (0.0676) & 0.329 (0.0583) & 0.325 (0.05)\\
\hline
Median & 0.446 (0.041) & 0.484 (0.078) & 0.449 (0.0937) & 0.444 (0.1012)\\
Q0.9 & 0.158 (0.1061) & 0.283 (0.0754) & 0.299 (0.148) & 0.409 (0.1113)\\
Q0.99 & 0.09 (0.0362) & 0.115 (0.0121) & 0.098 (0.0096) & 0.129 (0.0276)\\
Q0.999 & 0.303 (0.2673) & 0.239 (0.1036) & 0.322 (0.1004) & 0.441 (0.2547)\\
\hline
Threshold & 0 (0) & 0 (0) & 0 (0) & 0.075 (0.15)\\
\hline
\end{tabular}
\end{table}

\begin{table}[H]

\caption{Average false positive rates for changing number of extreme events.}
\centering
\begin{tabular}{l|l|l|l|l}
\hline
  & E =  1 & E =  2 & E =  3 & E =  4\\
\hline
ExLasso ($\gamma = 4$) & 0.002 (7e-04) & 0.01 (7e-04) & 0.003 (0.0013) & 0.001 (0.0011)\\
ExLasso ($\gamma = 6$) & 0.002 (8e-04) & 0.003 (0.0026) & 0.003 (0.0017) & 0.002 (0.0017)\\
\hline
ExSCAD 4th & 0.002 (7e-04) & 0.01 (7e-04) & 0.003 (0.0013) & 0.001 (0.0011)\\
ExSCAD 6th & 0.002 (8e-04) & 0.003 (0.0026) & 0.003 (0.0017) & 0.002 (0.0017)\\
\hline
ExMCP 4th & 0.002 (7e-04) & 0.01 (0.0014) & 0.003 (0.0013) & 0.001 (0.0011)\\
ExMCP 6th & 0.002 (8e-04) & 0.003 (0.0019) & 0.003 (0.0017) & 0.002 (0.0017)\\
\hline
Lasso & 0.009 (0) & 0.01 (7e-04) & 0.009 (0) & 0.009 (7e-04)\\
SCAD & 0.009 (0) & 0.01 (7e-04) & 0.009 (0) & 0.009 (7e-04)\\
MCP & 0.009 (0) & 0.01 (7e-04) & 0.009 (0) & 0.009 (7e-04)\\
\hline
Median & 0.011 (0) & 0.008 (0.0022) & 0.008 (0.0011) & 0.008 (0.0013)\\
Q0.9 & 0.01 (0.003) & 0.011 (0.0016) & 0.012 (0.0017) & 0.009 (0.0011)\\
Q0.99 & 0.018 (0.0059) & 0.01 (0.0013) & 0.012 (0.0013) & 0.011 (0.002)\\
Q0.999 & 0.013 (0.0078) & 0.01 (0.0017) & 0.01 (0.003) & 0.008 (0.0036)\\
\hline
Threshold & 0.01 (7e-04) & 0.01 (7e-04) & 0.009 (0) & 0.01 (7e-04)\\
\hline
\end{tabular}
\end{table}


\newpage

\noindent{\textit{Scenario 3: Changing Error Distribution}}

\begin{table}[H]

\caption{Average F-1 scores for changing residual distribution.}
\centering
\begin{tabular}{l|l|l|l|l}
\hline
  & $\beta$ =  0.33 & $\beta$ =  0.2 & $\beta$ =  0.125 & $\beta$ =  0.083\\
\hline
ExLasso ($\gamma = 4$) & 0.875 (0.05) & 0.8 (0.1155) & 0.625 (0.1258) & 0.275 (0.2217)\\
ExLasso ($\gamma = 6$) & 0.85 (0.0577) & 0.75 (0.1732) & 0.682 (0.1284) & 0.425 (0.15)\\
\hline
ExSCAD 4th & 0.875 (0.05) & 0.8 (0.1155) & 0.625 (0.1258) & 0.275 (0.2217)\\
ExSCAD 6th & 0.85 (0.0577) & 0.75 (0.1732) & 0.682 (0.1284) & 0.425 (0.15)\\
\hline
ExMCP 4th & 0.888 (0.0637) & 0.825 (0.0957) & 0.625 (0.1258) & 0.275 (0.2217)\\
ExMCP 6th & 0.864 (0.0474) & 0.788 (0.166) & 0.575 (0.0957) & 0.425 (0.15)\\
\hline
Lasso & 0.3 (0) & 0.2 (0.1414) & 0.262 (0.1103) & 0.175 (0.15)\\
SCAD & 0.3 (0) & 0.2 (0.1414) & 0.262 (0.1103) & 0.175 (0.15)\\
MCP & 0.3 (0) & 0.2 (0.1414) & 0.27 (0.1197) & 0.15 (0.1732)\\
\hline
Median & 0.529 (0.0626) & 0.338 (0.1134) & 0.46 (0.1078) & 0.403 (0.1414)\\
Q-0.9 & 0.185 (0.1239) & 0.122 (0.0465) & 0.121 (0.041) & 0.097 (0.0071)\\
Q-0.99 & 0.102 (0.0355) & 0.092 (0.0055) & 0.092 (0.0096) & 0.097 (0.0096)\\
Q-0.999 & 0.328 (0.2583) & 0.202 (0.1506) & 0.093 (0.0087) & 0.093 (0.0105)\\
\hline
Threshold & 0 (0) & 0.05 (0.1) & 0.073 (0.0994) & 0 (0)\\
\hline
\end{tabular}
\end{table}

\begin{table}[H]

\caption{Average true positive rates for changing residual distribution.}
\centering
\begin{tabular}{l|l|l|l|l}
\hline
  & $\beta$ =  0.33 & $\beta$ =  0.2 & $\beta$ =  0.125 & $\beta$ =  0.083\\
\hline
ExLasso ($\gamma = 4$) & 0.875 (0.05) & 0.8 (0.1155) & 0.625 (0.1258) & 0.275 (0.2217)\\
ExLasso ($\gamma = 6$) & 0.85 (0.0577) & 0.75 (0.1732) & 0.667 (0.1247) & 0.425 (0.15)\\
\hline
ExSCAD 4th & 0.875 (0.05) & 0.8 (0.1155) & 0.625 (0.1258) & 0.275 (0.2217)\\
ExSCAD 6th & 0.85 (0.0577) & 0.75 (0.1732) & 0.667 (0.1247) & 0.425 (0.15)\\
\hline
ExMCP 4th & 0.877 (0.0517) & 0.825 (0.0957) & 0.625 (0.1258) & 0.275 (0.2217)\\
ExMCP 6th & 0.855 (0.053) & 0.777 (0.1526) & 0.575 (0.0957) & 0.425 (0.15)\\
\hline
Lasso & 0.3 (0) & 0.2 (0.1414) & 0.252 (0.1013) & 0.175 (0.15)\\
SCAD & 0.3 (0) & 0.2 (0.1414) & 0.252 (0.1013) & 0.175 (0.15)\\
MCP & 0.3 (0) & 0.2 (0.1414) & 0.266 (0.1146) & 0.15 (0.1732)\\
\hline
Median & 0.446 (0.041) & 0.33 (0.0991) & 0.449 (0.0937) & 0.385 (0.1168)\\
Q-0.9 & 0.158 (0.1061) & 0.12 (0.0449) & 0.117 (0.034) & 0.094 (0.0136)\\
Q-0.99 & 0.09 (0.0362) & 0.086 (0.0099) & 0.086 (0.0176) & 0.096 (0.0199)\\
Q-0.999 & 0.303 (0.2673) & 0.207 (0.1615) & 0.087 (0.0163) & 0.088 (0.0184)\\
\hline
Threshold & 0 (0) & 0.05 (0.1) & 0.072 (0.1048) & 0 (0)\\
\hline
\end{tabular}
\end{table}

\begin{table}[H]

\caption{Average false positive rates for changing residual distribution.}
\centering
\begin{tabular}{l|l|l|l|l}
\hline
  & $\beta$ =  0.33 & $\beta$ =  0.2 & $\beta$ =  0.125 & $\beta$ =  0.083\\
\hline
ExLasso ($\gamma = 4$) & 0.002 (7e-04) & 0.003 (0.0016) & 0.005 (0.0017) & 0.01 (0.003)\\
ExLasso ($\gamma = 6$) & 0.002 (8e-04) & 0.003 (0.0023) & 0.005 (0.0017) & 0.008 (0.002)\\
\hline
ExSCAD 4th & 0.002 (7e-04) & 0.003 (0.0016) & 0.005 (0.0017) & 0.01 (0.003)\\
ExSCAD 6th & 0.002 (8e-04) & 0.003 (0.0023) & 0.005 (0.0017) & 0.008 (0.002)\\
\hline
ExMCP 4th & 0.002 (7e-04) & 0.002 (0.0013) & 0.005 (0.0017) & 0.01 (0.003)\\
ExMCP 6th & 0.002 (8e-04) & 0.003 (0.002) & 0.006 (0.0013) & 0.008 (0.002)\\
\hline
Lasso & 0.009 (0) & 0.012 (0.0039) & 0.011 (0.0016) & 0.011 (0.002)\\
SCAD & 0.009 (0) & 0.012 (0.0039) & 0.011 (0.0016) & 0.011 (0.002)\\
MCP & 0.009 (0) & 0.011 (0.0019) & 0.01 (0.0014) & 0.011 (0.0023)\\
\hline
Median & 0.011 (0) & 0.009 (0.0019) & 0.008 (0.0013) & 0.009 (8e-04)\\
Q-0.9 & 0.01 (0.003) & 0.012 (7e-04) & 0.014 (0.0023) & 0.013 (0.0013)\\
Q-0.99 & 0.018 (0.0059) & 0.015 (0.0017) & 0.015 (0.0029) & 0.013 (0.0026)\\
Q-0.999 & 0.013 (0.0078) & 0.011 (0.0033) & 0.015 (0.0026) & 0.015 (0.0034)\\
\hline
Threshold & 0.01 (7e-04) & 0.014 (0.0048) & 0.016 (0.0045) & 0.014 (0.0013)\\
\hline
\end{tabular}
\end{table}

%


\newpage

\noindent{\textit{Scenario 4: Changing Number of Dimensions}}

\begin{table}[H]

\caption{Average F-1 scores for changing number of dimensions.}
\centering
\begin{tabular}{l|l|l|l|l}
\hline
  & P =  750 & P =  1500 & P =  2250 & P =  3000\\
\hline
ExLasso ($\gamma = 4$) & 0.875 (0.05) & 0.75 (0.0577) & 0.827 (0.0848) & 0.627 (0.2906)\\
ExLasso ($\gamma = 6$) & 0.85 (0.0577) & 0.65 (0.1) & 0.642 (0.1962) & 0.55 (0.1)\\
\hline
ExSCAD 4th & 0.875 (0.05) & 0.75 (0.0577) & 0.827 (0.0848) & 0.627 (0.2906)\\
ExSCAD 6th & 0.85 (0.0577) & 0.65 (0.1) & 0.642 (0.1962) & 0.55 (0.1)\\
\hline
ExMCP 4th & 0.888 (0.0637) & 0.75 (0.0577) & 0.615 (0.2091) & 0.425 (0.2062)\\
ExMCP 6th & 0.864 (0.0474) & 0.664 (0.1218) & 0.521 (0.1279) & 0.6 (0)\\
\hline
Lasso & 0.3 (0) & 0 (0) & 0 (0) & 0 (0)\\
SCAD & 0.3 (0) & 0 (0) & 0 (0) & 0 (0)\\
MCP & 0.3 (0) & 0 (0) & 0 (0) & 0 (0)\\
\hline
Median & 0.529 (0.0626) & 0.249 (0.0671) & 0.247 (0.1579) & 0.103 (0.0269)\\
Q-0.9 & 0.185 (0.1239) & 0.117 (0.0553) & 0.102 (0.0119) & 0.103 (0.0269)\\
Q-0.99 & 0.102 (0.0355) & 0.1 (0.0041) & 0.089 (0.0051) & 0.095 (0.0037)\\
Q-0.999 & 0.328 (0.2583) & 0.117 (0.0495) & 0.279 (0.2265) & 0.089 (0.0051)\\
\hline
Threshold & 0 (0) & 0 (0) & 0 (0) & 0 (0)\\
\hline
\end{tabular}
\end{table}

\begin{table}[H]

\caption{Average true positive rates for changing number of dimensions.}
\centering
\begin{tabular}{l|l|l|l|l}
\hline
  & P =  750 & P =  1500 & P =  2250 & P =  3000\\
\hline
ExLasso ($\gamma = 4$) & 0.875 (0.05) & 0.75 (0.0577) & 0.752 (0.0346) & 0.542 (0.2378)\\
ExLasso ($\gamma = 6$) & 0.85 (0.0577) & 0.65 (0.1) & 0.617 (0.1607) & 0.55 (0.1)\\
\hline
ExSCAD 4th & 0.875 (0.05) & 0.75 (0.0577) & 0.752 (0.0346) & 0.542 (0.2378)\\
ExSCAD 6th & 0.85 (0.0577) & 0.65 (0.1) & 0.617 (0.1607) & 0.55 (0.1)\\
\hline
ExMCP 4th & 0.877 (0.0517) & 0.75 (0.0577) & 0.663 (0.2358) & 0.425 (0.2062)\\
ExMCP 6th & 0.855 (0.053) & 0.672 (0.0713) & 0.544 (0.1423) & 0.6 (0)\\
\hline
Lasso & 0.3 (0) & 0 (0) & 0 (0) & 0 (0)\\
SCAD & 0.3 (0) & 0 (0) & 0 (0) & 0 (0)\\
MCP & 0.3 (0) & 0 (0) & 0 (0) & 0 (0)\\
\hline
Median & 0.446 (0.041) & 0.229 (0.0473) & 0.247 (0.1426) & 0.124 (0.0845)\\
Q-0.9 & 0.158 (0.1061) & 0.111 (0.0596) & 0.107 (0.0266) & 0.124 (0.0845)\\
Q-0.99 & 0.09 (0.0362) & 0.101 (0.0083) & 0.081 (0.0084) & 0.091 (0.0068)\\
Q-0.999 & 0.303 (0.2673) & 0.112 (0.0494) & 0.267 (0.2336) & 0.081 (0.0084)\\
\hline
Threshold & 0 (0) & 0 (0) & 0 (0) & 0 (0)\\
\hline
\end{tabular}
\end{table}

\begin{table}[H]

\caption{Average false positive rates for changing number of dimensions.}
\centering
\begin{tabular}{l|l|l|l|l}
\hline
  & P =  750 & P =  1500 & P =  2250 & P =  3000\\
\hline
ExLasso ($\gamma = 4$) & 0.002 (7e-04) & 0.003 (8e-04) & 0.004 (0) & 0.008 (0.0028)\\
ExLasso ($\gamma = 6$) & 0.002 (8e-04) & 0.005 (0.0014) & 0.005 (0.0019) & 0.006 (0.0014)\\
\hline
ExSCAD 4th & 0.002 (7e-04) & 0.003 (8e-04) & 0.004 (0) & 0.008 (0.0028)\\
ExSCAD 6th & 0.002 (8e-04) & 0.005 (0.0014) & 0.005 (0.0019) & 0.006 (0.0014)\\
\hline
ExMCP 4th & 0.002 (7e-04) & 0.003 (8e-04) & 0.004 (0.0029) & 0.008 (0.0028)\\
ExMCP 6th & 0.002 (8e-04) & 0.004 (0.0013) & 0.006 (0.002) & 0.005 (0)\\
\hline
Lasso & 0.009 (0) & 0.015 (0.002) & 0.014 (0) & 0.015 (0.002)\\
SCAD & 0.009 (0) & 0.015 (0.002) & 0.014 (0) & 0.015 (0.002)\\
MCP & 0.009 (0) & 0.014 (0.0022) & 0.013 (0.0014) & 0.014 (0.0014)\\
\hline
Median & 0.011 (0) & 0.012 (0.0011) & 0.01 (0.0026) & 0.012 (0.0058)\\
Q-0.9 & 0.01 (0.003) & 0.014 (0.0028) & 0.013 (0.0028) & 0.015 (0.0029)\\
Q-0.99 & 0.018 (0.0059) & 0.012 (0.0011) & 0.016 (0.0017) & 0.014 (0.0011)\\
Q-0.999 & 0.013 (0.0078) & 0.014 (0.0028) & 0.012 (0.0051) & 0.016 (0.0017)\\
\hline
Threshold & 0.01 (7e-04) & 0.021 (7e-04) & 0.02 (0.0098) & 0.016 (0.0052)\\
\hline
\end{tabular}
\end{table}

\subsection{Mixture Model Simulation Study}

\noindent{\textit{Scenario 1: Changing Magnitude of Extreme Values of Response Variable}}

\begin{table}[H]

\caption{Average F-1 scores for changing magnitude of extreme value magnitude.}
\centering
\begin{tabular}{l|l|l|l|l}
\hline
  & $\tau$ =  6 & $\tau$ =  7 & $\tau$ =  9 & $\tau$ =  50\\
\hline
ExLasso ($\gamma = 4$) & 0.128 (0.0986) & 0.175 (0.1708) & 0.9 (0.0816) & 1 (0)\\
ExLasso ($\gamma = 6$) & 0.259 (0.3143) & 0.757 (0.0963) & 1 (0) & 1 (0)\\
\hline
ExSCAD 4th & 0.128 (0.0986) & 0.175 (0.1708) & 0.9 (0.0816) & 1 (0)\\
ExSCAD 6th & 0.259 (0.3143) & 0.757 (0.0963) & 1 (0) & 1 (0)\\
\hline
ExMCP 4th & 0.028 (0.0556) & 0.106 (0.1222) & 0.82 (0.0688) & 0.972 (0.0556)\\
ExMCP 6th & 0.105 (0.2105) & 0.653 (0.1974) & 0.946 (0.0454) & 1 (0)\\
\hline
Lasso & 0 (0) & 0.075 (0.15) & 0 (0) & 1 (0)\\
SCAD & 0 (0) & 0.075 (0.15) & 0 (0) & 1 (0)\\
MCP & 0 (0) & 0.075 (0.15) & 0 (0) & 1 (0)\\
\hline
Median & 0.094 (0.0022) & 0.185 (0.1797) & 0.095 (0) & 0.095 (0)\\
Q0.9 & 0.094 (0.0022) & 0.14 (0.0887) & 0.095 (0) & 0.095 (0)\\
Q0.99 & 0.179 (0.0599) & 0.098 (0.0193) & 0.312 (0.1434) & 0.739 (0.1504)\\
Q0.999 & 0.348 (0.0986) & 0.369 (0.1994) & 0.394 (0.1643) & 0.474 (0.1721)\\
\hline
Threshold & 0 (0) & 0.123 (0.1798) & 0.384 (0.392) & 0.977 (0.0455)\\
\hline
\end{tabular}
\end{table}

\begin{table}[H]

\caption{Average true positive rates for changing magnitude of extreme value magnitude.}
\centering
\begin{tabular}{l|l|l|l|l}
\hline
  & $\tau$ =  6 & $\tau$ =  7 & $\tau$ =  9 & $\tau$ =  50\\
\hline
ExLasso ($\gamma = 4$) & 0.131 (0.102) & 0.175 (0.1708) & 0.9 (0.0816) & 1 (0)\\
ExLasso ($\gamma = 6$) & 0.246 (0.2936) & 0.742 (0.1067) & 1 (0) & 1 (0)\\
\hline
ExSCAD 4th & 0.131 (0.102) & 0.175 (0.1708) & 0.9 (0.0816) & 1 (0)\\
ExSCAD 6th & 0.246 (0.2936) & 0.742 (0.1067) & 1 (0) & 1 (0)\\
\hline
ExMCP 4th & 0.031 (0.0625) & 0.112 (0.1315) & 0.842 (0.0618) & 1 (0)\\
ExMCP 6th & 0.111 (0.2222) & 0.686 (0.1878) & 1 (0) & 1 (0)\\
\hline
Lasso & 0 (0) & 0.075 (0.15) & 0 (0) & 1 (0)\\
SCAD & 0 (0) & 0.075 (0.15) & 0 (0) & 1 (0)\\
MCP & 0 (0) & 0.075 (0.15) & 0 (0) & 1 (0)\\
\hline
Median & 0.089 (0.0038) & 0.172 (0.1629) & 0.091 (0) & 0.091 (0)\\
Q0.9 & 0.089 (0.0038) & 0.131 (0.0795) & 0.091 (0) & 0.091 (0)\\
Q0.99 & 0.166 (0.0587) & 0.1 (0.0322) & 0.303 (0.1415) & 0.671 (0.1675)\\
Q0.999 & 0.389 (0.1361) & 0.346 (0.188) & 0.373 (0.1713) & 0.426 (0.1602)\\
\hline
Threshold & 0 (0) & 0.122 (0.1714) & 0.353 (0.3505) & 0.958 (0.0833)\\
\hline
\end{tabular}
\end{table}

\begin{table}[H]

\caption{Average false positive rates for changing magnitude of extreme value magnitude.}
\centering
\begin{tabular}{l|l|l|l|l}
\hline
  & $\tau$ =  6 & $\tau$ =  7 & $\tau$ =  9 & $\tau$ =  50\\
\hline
ExLasso ($\gamma = 4$) & 0.011 (0.0013) & 0.011 (0.0023) & 0.001 (0.0011) & 0 (0)\\
ExLasso ($\gamma = 6$) & 0.01 (0.0028) & 0.004 (0.0017) & 0 (0) & 0 (0)\\
\hline
ExSCAD 4th & 0.011 (0.0013) & 0.011 (0.0023) & 0.001 (0.0011) & 0 (0)\\
ExSCAD 6th & 0.01 (0.0028) & 0.004 (0.0017) & 0 (0) & 0 (0)\\
\hline
ExMCP 4th & 0.01 (8e-04) & 0.01 (0.0014) & 0.002 (8e-04) & 0 (0)\\
ExMCP 6th & 0.01 (0.0023) & 0.004 (0.002) & 0 (0) & 0 (0)\\
\hline
Lasso & 0.013 (7e-04) & 0.012 (0.002) & 0.014 (0) & 0 (0)\\
SCAD & 0.013 (7e-04) & 0.012 (0.002) & 0.014 (0) & 0 (0)\\
MCP & 0.013 (7e-04) & 0.012 (0.0019) & 0.013 (0.0014) & 0 (0)\\
\hline
Median & 0.014 (7e-04) & 0.012 (0.002) & 0.014 (0) & 0.014 (0)\\
Q0.9 & 0.014 (8e-04) & 0.014 (8e-04) & 0.014 (0) & 0.014 (7e-04)\\
Q0.99 & 0.014 (0.004) & 0.014 (0.007) & 0.01 (0.0026) & 0.006 (0.0032)\\
Q0.999 & 0.007 (0.0032) & 0.011 (0.0052) & 0.01 (0.0036) & 0.01 (0.0051)\\
\hline
Threshold & 0.009 (0.0023) & 0.011 (0.0019) & 0.008 (0.0034) & 0.001 (0.0014)\\
\hline
\end{tabular}
\end{table}

%


\newpage

\noindent{\textit{Scenario 2: Changing Number of Extreme Events in Response}}

\begin{table}[H]

\caption{Average F-1 scores for changing number of extreme events.}
\centering
\begin{tabular}{l|l|l|l|l}
\hline
  & E =  1 & E =  2 & E =  3 & E =  4\\
\hline
ExLasso ($\gamma = 4$) & 0.128 (0.0986) & 0.313 (0.1514) & 0.632 (0.1489) & 0.836 (0.0473)\\
ExLasso ($\gamma = 6$) & 0.259 (0.3143) & 0.52 (0.0869) & 0.795 (0.1527) & 0.908 (0.0789)\\
\hline
ExSCAD 4th & 0.128 (0.0986) & 0.313 (0.1514) & 0.632 (0.1489) & 0.836 (0.0473)\\
ExSCAD 6th & 0.259 (0.3143) & 0.52 (0.0869) & 0.795 (0.1527) & 0.908 (0.0789)\\
\hline
ExMCP 4th & 0.028 (0.0556) & 0.164 (0.136) & 0.559 (0.2802) & 0.743 (0.1306)\\
ExMCP 6th & 0.105 (0.2105) & 0.239 (0.2046) & 0.697 (0.2623) & 0.83 (0.0989)\\
\hline
Lasso & 0 (0) & 0 (0) & 0 (0) & 0 (0)\\
SCAD & 0 (0) & 0 (0) & 0 (0) & 0 (0)\\
MCP & 0 (0) & 0 (0) & 0 (0) & 0 (0)\\
\hline
Median & 0.094 (0.0022) & 0.094 (0.0022) & 0.095 (0) & 0.095 (0)\\
Q0.9 & 0.094 (0.0022) & 0.094 (0.0022) & 0.095 (0) & 0.095 (0)\\
Q0.99 & 0.179 (0.0599) & 0.421 (0.1032) & 0.604 (0.093) & 0.65 (0.1238)\\
Q0.999 & 0.348 (0.0986) & 0.358 (0.0519) & 0.45 (0.1935) & 0.474 (0.1154)\\
\hline
Threshold & 0 (0) & 0.229 (0.1455) & 0.596 (0.1489) & 0.758 (0.1173)\\
\hline
\end{tabular}
\end{table}

\begin{table}[H]

\caption{Average true positive rates for changing number of extreme events.}
\centering
\begin{tabular}{l|l|l|l|l}
\hline
  & E =  1 & E =  2 & E =  3 & E =  4\\
\hline
ExLasso ($\gamma = 4$) & 0.131 (0.102) & 0.328 (0.1627) & 0.667 (0.1571) & 0.847 (0.0547)\\
ExLasso ($\gamma = 6$) & 0.246 (0.2936) & 0.542 (0.0949) & 0.817 (0.1599) & 0.944 (0.0642)\\
\hline
ExSCAD 4th & 0.131 (0.102) & 0.328 (0.1627) & 0.667 (0.1571) & 0.847 (0.0547)\\
ExSCAD 6th & 0.246 (0.2936) & 0.542 (0.0949) & 0.817 (0.1599) & 0.944 (0.0642)\\
\hline
ExMCP 4th & 0.031 (0.0625) & 0.182 (0.144) & 0.599 (0.2817) & 0.837 (0.0834)\\
ExMCP 6th & 0.111 (0.2222) & 0.259 (0.2049) & 0.761 (0.2223) & 0.937 (0.0745)\\
\hline
Lasso & 0 (0) & 0 (0) & 0 (0) & 0 (0)\\
SCAD & 0 (0) & 0 (0) & 0 (0) & 0 (0)\\
MCP & 0 (0) & 0 (0) & 0 (0) & 0 (0)\\
\hline
Median & 0.089 (0.0038) & 0.089 (0.0038) & 0.091 (0) & 0.091 (0)\\
Q0.9 & 0.089 (0.0038) & 0.089 (0.0038) & 0.091 (0) & 0.091 (0)\\
Q0.99 & 0.166 (0.0587) & 0.397 (0.0874) & 0.588 (0.1032) & 0.691 (0.0929)\\
Q0.999 & 0.389 (0.1361) & 0.369 (0.0525) & 0.479 (0.2206) & 0.479 (0.1158)\\
\hline
Threshold & 0 (0) & 0.234 (0.1401) & 0.653 (0.1768) & 0.767 (0.1054)\\
\hline
\end{tabular}
\end{table}

\begin{table}[H]

\caption{Average false positive rates for changing number of extreme events.}
\centering
\begin{tabular}{l|l|l|l|l}
\hline
  & E =  1 & E =  2 & E =  3 & E =  4\\
\hline
ExLasso ($\gamma = 4$) & 0.011 (0.0013) & 0.008 (0.0023) & 0.004 (0.0019) & 0.002 (8e-04)\\
ExLasso ($\gamma = 6$) & 0.01 (0.0028) & 0.006 (0.0013) & 0.002 (0.002) & 0.001 (8e-04)\\
\hline
ExSCAD 4th & 0.011 (0.0013) & 0.008 (0.0023) & 0.004 (0.0019) & 0.002 (8e-04)\\
ExSCAD 6th & 0.01 (0.0028) & 0.006 (0.0013) & 0.002 (0.002) & 0.001 (8e-04)\\
\hline
ExMCP 4th & 0.01 (8e-04) & 0.008 (7e-04) & 0.004 (0.0028) & 0.002 (7e-04)\\
ExMCP 6th & 0.01 (0.0023) & 0.008 (0.0013) & 0.002 (0.002) & 0.001 (8e-04)\\
\hline
Lasso & 0.013 (7e-04) & 0.013 (7e-04) & 0.013 (7e-04) & 0.013 (7e-04)\\
SCAD & 0.013 (7e-04) & 0.013 (7e-04) & 0.013 (7e-04) & 0.013 (7e-04)\\
MCP & 0.013 (7e-04) & 0.012 (0.0013) & 0.012 (0.0019) & 0.012 (0.0019)\\
\hline
Median & 0.014 (7e-04) & 0.014 (7e-04) & 0.014 (0) & 0.014 (0)\\
Q0.9 & 0.014 (8e-04) & 0.014 (0) & 0.014 (0.0014) & 0.014 (8e-04)\\
Q0.99 & 0.014 (0.004) & 0.009 (0.0013) & 0.006 (0.0023) & 0.004 (0.0013)\\
Q0.999 & 0.007 (0.0032) & 0.008 (0.0011) & 0.007 (0.0038) & 0.007 (0.0026)\\
\hline
Threshold & 0.009 (0.0023) & 0.009 (0.0011) & 0.004 (0.0022) & 0.003 (0.0013)\\
\hline
\end{tabular}
\end{table}

%


\newpage

\noindent{\textit{Scenario 3: Changing Error Distribution}}

\begin{table}[H]

\caption{Average F-1 scores for changing residual distribution.}
\centering
\begin{tabular}{l|l|l|l|l}
\hline
  & $\beta$  =  0.33 & $\beta$  =  0.2 & $\beta$  =  0.166 & $\beta$  =  0.125\\
\hline
ExLasso ($\gamma = 4$) & 0.9 (0.0816) & 0.875 (0.1258) & 0.816 (0.0526) & 0.278 (0.3587)\\
ExLasso ($\gamma = 6$) & 1 (0) & 1 (0) & 0.582 (0.2852) & 0.184 (0.217)\\
\hline
ExSCAD 4th & 0.9 (0.0816) & 0.875 (0.1258) & 0.816 (0.0526) & 0.278 (0.3587)\\
ExSCAD 6th & 1 (0) & 1 (0) & 0.582 (0.2852) & 0.184 (0.217)\\
\hline
ExMCP 4th & 0.82 (0.0688) & 0.818 (0.0819) & 0.735 (0.1126) & 0.288 (0.3796)\\
ExMCP 6th & 0.946 (0.0454) & 0.917 (0.0556) & 0.693 (0.1714) & 0.144 (0.1744)\\
\hline
Lasso & 0 (0) & 0 (0) & 0 (0) & 0 (0)\\
SCAD & 0 (0) & 0 (0) & 0 (0) & 0 (0)\\
MCP & 0 (0) & 0 (0) & 0 (0) & 0 (0)\\
\hline
Median & 0.095 (0) & 0.095 (0) & 0.095 (0) & 0.094 (0.0022)\\
Q-0.9 & 0.095 (0) & 0.095 (0) & 0.095 (0) & 0.094 (0.0022)\\
Q-0.99 & 0.312 (0.1434) & 0.14 (0.0494) & 0.249 (0.0897) & 0.24 (0.0465)\\
Q-0.999 & 0.394 (0.1643) & 0.299 (0.0897) & 0.19 (0.0186) & 0.115 (0.0505)\\
\hline
Threshold & 0.384 (0.392) & 0.05 (0.1) & 0.64 (0.0773) & 0.508 (0.2058)\\
\hline
\end{tabular}
\end{table}

\begin{table}[H]

\caption{Average true positive rates for changing residual distribution.}
\centering
\begin{tabular}{l|l|l|l|l}
\hline
  & $\beta$  =  0.33 & $\beta$  =  0.2 & $\beta$  =  0.166 & $\beta$  =  0.125\\
\hline
ExLasso ($\gamma = 4$) & 0.9 (0.0816) & 0.875 (0.1258) & 0.861 (0.0556) & 0.281 (0.358)\\
ExLasso ($\gamma = 6$) & 1 (0) & 1 (0) & 0.589 (0.2837) & 0.194 (0.2291)\\
\hline
ExSCAD 4th & 0.9 (0.0816) & 0.875 (0.1258) & 0.861 (0.0556) & 0.281 (0.358)\\
ExSCAD 6th & 1 (0) & 1 (0) & 0.589 (0.2837) & 0.194 (0.2291)\\
\hline
ExMCP 4th & 0.842 (0.0618) & 0.869 (0.1245) & 0.893 (0.1368) & 0.307 (0.3857)\\
ExMCP 6th & 1 (0) & 1 (0) & 0.821 (0.1798) & 0.17 (0.209)\\
\hline
Lasso & 0 (0) & 0 (0) & 0 (0) & 0 (0)\\
SCAD & 0 (0) & 0 (0) & 0 (0) & 0 (0)\\
MCP & 0 (0) & 0 (0) & 0 (0) & 0 (0)\\
\hline
Median & 0.091 (0) & 0.091 (0) & 0.091 (0) & 0.089 (0.0038)\\
Q-0.9 & 0.091 (0) & 0.091 (0) & 0.091 (0) & 0.089 (0.0038)\\
Q-0.99 & 0.303 (0.1415) & 0.133 (0.0438) & 0.251 (0.0829) & 0.232 (0.0391)\\
Q-0.999 & 0.373 (0.1713) & 0.277 (0.0854) & 0.182 (0.0343) & 0.107 (0.0505)\\
\hline
Threshold & 0.353 (0.3505) & 0.05 (0.1) & 0.689 (0.092) & 0.517 (0.2134)\\
\hline
\end{tabular}
\end{table}

\begin{table}[H]

\caption{Average false positive rates for changing residual distribution.}
\centering
\begin{tabular}{l|l|l|l|l}
\hline
  & $\beta$  =  0.33 & $\beta$  =  0.2 & $\beta$  =  0.166 & $\beta$  =  0.125\\
\hline
ExLasso ($\gamma = 4$) & 0.001 (0.0011) & 0.002 (0.0017) & 0.002 (7e-04) & 0.009 (0.0045)\\
ExLasso ($\gamma = 6$) & 0 (0) & 0 (0) & 0.005 (0.0038) & 0.01 (0.0032)\\
\hline
ExSCAD 4th & 0.001 (0.0011) & 0.002 (0.0017) & 0.002 (7e-04) & 0.009 (0.0045)\\
ExSCAD 6th & 0 (0) & 0 (0) & 0.005 (0.0038) & 0.01 (0.0032)\\
\hline
ExMCP 4th & 0.002 (8e-04) & 0.002 (0.0017) & 0.001 (0.0013) & 0.007 (0.0033)\\
ExMCP 6th & 0 (0) & 0 (0) & 0.002 (0.0017) & 0.009 (0.0026)\\
\hline
Lasso & 0.014 (0) & 0.014 (0) & 0.014 (0) & 0.013 (7e-04)\\
SCAD & 0.014 (0) & 0.014 (0) & 0.014 (0) & 0.013 (7e-04)\\
MCP & 0.013 (0.0014) & 0.013 (7e-04) & 0.012 (0.0016) & 0.012 (0.0019)\\
\hline
Median & 0.014 (0) & 0.014 (0) & 0.014 (0) & 0.014 (7e-04)\\
Q-0.9 & 0.014 (0) & 0.015 (7e-04) & 0.013 (0.0014) & 0.013 (0.0023)\\
Q-0.99 & 0.01 (0.0026) & 0.013 (0.002) & 0.01 (0.0023) & 0.011 (0.0013)\\
Q-0.999 & 0.01 (0.0036) & 0.011 (0.0017) & 0.012 (0.0028) & 0.015 (0.002)\\
\hline
Threshold & 0.008 (0.0034) & 0.011 (7e-04) & 0.004 (0.0013) & 0.006 (0.003)\\
\hline
\end{tabular}
\end{table}

%


\newpage

\noindent{\textit{Scenario 4: Changing Number of Dimensions}}

\begin{table}[H]

\caption{Average F-1 scores for changing number of dimensions.}
\centering
\begin{tabular}{l|l|l|l|l}
\hline
  & P =  750 & P =  1500 & P =  2250 & P =  3000\\
\hline
ExLasso ($\gamma = 4$) & 0.9 (0.0816) & 0.816 (0.0526) & 0.922 (0.0673) & 0.838 (0.0062)\\
ExLasso ($\gamma = 6$) & 1 (0) & 0.947 (0) & 0.961 (0.0263) & 0.947 (0)\\
\hline
ExSCAD 4th & 0.9 (0.0816) & 0.816 (0.0526) & 0.922 (0.0673) & 0.838 (0.0062)\\
ExSCAD 6th & 1 (0) & 0.947 (0) & 0.961 (0.0263) & 0.947 (0)\\
\hline
ExMCP 4th & 0.82 (0.0688) & 0.782 (0.0708) & 0.765 (0.0679) & 0.683 (0.134)\\
ExMCP 6th & 0.946 (0.0454) & 0.854 (0.0619) & 0.961 (0.0263) & 0.885 (0.0876)\\
\hline
Lasso & 0 (0) & 0 (0) & 0 (0) & 0 (0)\\
SCAD & 0 (0) & 0 (0) & 0 (0) & 0 (0)\\
MCP & 0 (0) & 0 (0) & 0 (0) & 0 (0)\\
\hline
Median & 0.095 (0) & 0.095 (0) & 0.095 (0) & 0.095 (0)\\
Q-0.9 & 0.095 (0) & 0.095 (0) & 0.095 (0) & 0.095 (0)\\
Q-0.99 & 0.312 (0.1434) & 0.238 (0.0794) & 0.189 (0.0744) & 0.093 (0.0262)\\
Q-0.999 & 0.394 (0.1643) & 0.39 (0.0962) & 0.501 (0.0663) & 0.577 (0.2264)\\
\hline
Threshold & 0.384 (0.392) & 0.32 (0.1879) & 0.316 (0.087) & 0 (0)\\
\hline
\end{tabular}
\end{table}

\begin{table}[H]

\caption{Average true positive rates for changing number of dimensions.}
\centering
\begin{tabular}{l|l|l|l|l}
\hline
  & P =  750 & P =  1500 & P =  2250 & P =  3000\\
\hline
ExLasso ($\gamma = 4$) & 0.9 (0.0816) & 0.861 (0.0556) & 0.947 (0.0611) & 0.802 (0.1235)\\
ExLasso ($\gamma = 6$) & 1 (0) & 1 (0) & 1 (0) & 1 (0)\\
\hline
ExSCAD 4th & 0.9 (0.0816) & 0.861 (0.0556) & 0.947 (0.0611) & 0.802 (0.1235)\\
ExSCAD 6th & 1 (0) & 1 (0) & 1 (0) & 1 (0)\\
\hline
ExMCP 4th & 0.842 (0.0618) & 0.853 (0.0524) & 0.929 (0.0825) & 0.795 (0.1136)\\
ExMCP 6th & 1 (0) & 1 (0) & 1 (0) & 1 (0)\\
\hline
Lasso & 0 (0) & 0 (0) & 0 (0) & 0 (0)\\
SCAD & 0 (0) & 0 (0) & 0 (0) & 0 (0)\\
MCP & 0 (0) & 0 (0) & 0 (0) & 0 (0)\\
\hline
Median & 0.091 (0) & 0.091 (0) & 0.091 (0) & 0.091 (0)\\
Q-0.9 & 0.091 (0) & 0.091 (0) & 0.091 (0) & 0.091 (0)\\
Q-0.99 & 0.303 (0.1415) & 0.207 (0.0762) & 0.18 (0.0684) & 0.092 (0.0468)\\
Q-0.999 & 0.373 (0.1713) & 0.371 (0.1244) & 0.486 (0.0278) & 0.567 (0.2974)\\
\hline
Threshold & 0.353 (0.3505) & 0.344 (0.1929) & 0.335 (0.0958) & 0 (0)\\
\hline
\end{tabular}
\end{table}

\begin{table}[H]

\caption{Average false positive rates for changing number of dimensions.}
\centering
\begin{tabular}{l|l|l|l|l}
\hline
  & P =  750 & P =  1500 & P =  2250 & P =  3000\\
\hline
ExLasso ($\gamma = 4$) & 0.001 (0.0011) & 0.002 (7e-04) & 0.001 (8e-04) & 0.003 (0.0029)\\
ExLasso ($\gamma = 6$) & 0 (0) & 0 (0) & 0 (0) & 0 (0)\\
\hline
ExSCAD 4th & 0.001 (0.0011) & 0.002 (7e-04) & 0.001 (8e-04) & 0.003 (0.0029)\\
ExSCAD 6th & 0 (0) & 0 (0) & 0 (0) & 0 (0)\\
\hline
ExMCP 4th & 0.002 (8e-04) & 0.002 (7e-04) & 0.001 (8e-04) & 0.002 (0.001)\\
ExMCP 6th & 0 (0) & 0 (0) & 0 (0) & 0 (0)\\
\hline
Lasso & 0.014 (0) & 0.013 (8e-04) & 0.013 (8e-04) & 0.013 (0.001)\\
SCAD & 0.014 (0) & 0.013 (8e-04) & 0.013 (8e-04) & 0.013 (0.001)\\
MCP & 0.013 (0.0014) & 0.011 (0.0013) & 0.012 (0.0017) & 0.013 (0.001)\\
\hline
Median & 0.014 (0) & 0.014 (0) & 0.014 (0) & 0.014 (0)\\
Q-0.9 & 0.014 (0) & 0.017 (0.0013) & 0.015 (0.0026) & 0.014 (0.001)\\
Q-0.99 & 0.01 (0.0026) & 0.017 (0.0078) & 0.012 (0.0019) & 0.016 (0.0086)\\
Q-0.999 & 0.01 (0.0036) & 0.01 (0.0045) & 0.007 (0.0014) & 0.007 (0.0067)\\
\hline
Threshold & 0.008 (0.0034) & 0.007 (0.0017) & 0.008 (0.0016) & 0.012 (0.0019)\\
\hline
\end{tabular}
\end{table}

\newpage

\section*{Acknowledgements}

The authors acknowledge support from NSF DMS-1554821 and NSF NeuroNex-1707400. We also thank Dr. Michael Weylandt for providing useful discussions.

\newpage

\bibliography{bibb}
\end{document}